\documentclass[12pt,a4paper]{article}
\usepackage[super,sort&compress,comma]{natbib} 
\usepackage[version=3]{mhchem}
\usepackage[left=1.5cm, right=1.5cm, top=1.785cm, bottom=2.0cm]{geometry}
\usepackage{balance}
\usepackage{mathptmx}
\usepackage{sectsty}
\usepackage{graphicx} 
\usepackage{lastpage}
\usepackage{url}
\usepackage[format=plain,justification=justified,singlelinecheck=false,font={stretch=1.125,small,sf},labelfont=bf,labelsep=space]{caption}
\usepackage{float}
\usepackage{fancyhdr}
\usepackage{fnpos}
\usepackage[english]{babel}
\addto{\captionsenglish}{%

}
\usepackage{array}
\usepackage{droidsans}
\usepackage{charter}
\usepackage[T1]{fontenc}
\usepackage[usenames,dvipsnames]{xcolor}
\usepackage{setspace}
\usepackage[compact]{titlesec}
\usepackage{hyperref}

\usepackage{epstopdf}

\definecolor{cream}{RGB}{222,217,201}

\begin{document}

\pagestyle{fancy}
\thispagestyle{plain}
\fancypagestyle{plain}{
\renewcommand{\headrulewidth}{0pt}
}


\renewcommand{\thefootnote}{\fnsymbol{footnote}}
\renewcommand\footnoterule{\vspace*{1pt}%
\color{cream}\hrule width 3.5in height 0.4pt \color{black}\vspace*{5pt}} 
\setcounter{secnumdepth}{5}

\makeatletter 
\renewcommand\@biblabel[1]{#1}            
\renewcommand\@makefntext[1]%
{\noindent\makebox[0pt][r]{\@thefnmark\,}#1}
\makeatother 
\renewcommand{\figurename}{\small{Fig.}~}
\sectionfont{\sffamily\Large}
\subsectionfont{\normalsize}
\subsubsectionfont{\bf}
\setstretch{1.125} 
\setlength{\skip\footins}{0.8cm}
\setlength{\footnotesep}{0.25cm}
\setlength{\jot}{10pt}
\titlespacing*{\section}{0pt}{4pt}{4pt}
\titlespacing*{\subsection}{0pt}{15pt}{1pt}

\renewcommand{\headrulewidth}{0pt} 
\renewcommand{\footrulewidth}{0pt}
\setlength{\arrayrulewidth}{1pt}
\setlength{\columnsep}{6.5mm}
\setlength\bibsep{1pt}

\makeatletter 
\newlength{\figrulesep} 
\setlength{\figrulesep}{0.5\textfloatsep} 

\newcommand{\topfigrule}{\vspace*{-1pt}%
\noindent{\color{cream}\rule[-\figrulesep]{\columnwidth}{1.5pt}} }

\newcommand{\botfigrule}{\vspace*{-2pt}%
\noindent{\color{cream}\rule[\figrulesep]{\columnwidth}{1.5pt}} }

\newcommand{\dblfigrule}{\vspace*{-1pt}%
\noindent{\color{cream}\rule[-\figrulesep]{\textwidth}{1.5pt}} }

\makeatother


\begin{center}

{\LARGE Valence Ionization Of Water Clusters Formed Inside Helium Nanodroplets \par}

\vspace{0.8cm}

{\large
Subhendu De$^{1,6}$,
Sivarama Krishnan$^{1}$,
Niklas Sheel$^{2}$,
Keshav Sishodia$^{1}$,
Robert Richter$^{5}$,
Marcel Mudrich$^{2}$,
Florent Calvo$^{3}$,
Ltaief Ben Ltaief$^{2,*}$
\par}

\vspace{0.5cm}

{\small
$^{1}$ Quantum Center of Excellence for Diamond and Emergent Materials and Department of Physics,
Indian Institute of Technology Madras, Chennai 600036, India\\
$^{2}$ Department of Physics and Astronomy, Aarhus University, 8000 Aarhus C, Denmark\\
$^{3}$ Université Grenoble Alpes, CNRS, LIPhy, 38000 Grenoble, France\\
$^{5}$ Elettra-Sincrotrone Trieste, 34149 Basovizza, Italy\\
$^{6}$ Laboratoire de Chimie Physique--Matière et Rayonnement (LCPMR),
UMR 7614, CNRS, Sorbonne Université, 75005 Paris, France
}

\vspace{0.8cm}

\begin{minipage}{0.9\textwidth}
\small
\textbf{Abstract.}
The ionization mechanisms of small $\mathrm{H_2O}$/$\mathrm{D_2O}$ clusters embedded in helium nanodroplets (HNDs) irradiated with extreme ultraviolet photons of energy $h\nu=21.6$~eV are investigated using Penning ionization electron-ion coincidence spectroscopy. Both protonated $\mathrm{(H_2O)}_{n-1}\mathrm{H}^{+}$/$\mathrm{(D_2O)}_{n-1}\mathrm{D}^{+}$ ($n=3$--6) and unprotonated $\mathrm{(H_2O)}_{n}^{+}$/$\mathrm{(D_2O)}_{n}^{+}$ ($n=2$--5) cluster ions were observed. Penning ionization electron spectra measured in coincidence with water cluster ions emitted from water clusters doped into both large and small HNDs are analyzed and compared with photoelectron-photoion coincidence spectra measured for free water clusters at $h\nu=20.6$~eV. The results reveal suppression of fragmentation inside HNDs and stabilization of intact cluster ions. Quantum chemical calculations support the coexistence of proton-transferred and hemibonded conformers under the cryogenic conditions of helium nanodroplets.
\end{minipage}

\end{center}

\vspace{1cm}


\footnotetext{\textit{$^{1}$~Quantum Center of Excellence for Diamond and Emergent Materials and Department of Physics, Indian Institute of Technology Madras, Chennai 600036, India.}}

\footnotetext{\textit{$^{2}$~Department of Physics and Astronomy, Aarhus University, 8000 Aarhus C, Denmark.}}
\footnotetext{\textit{$^{3}$~Université Grenoble Alpes, CNRS, LIPhy, 38000 Grenoble, France.}}
\footnotetext{\textit{$^{5}$~Elettra-Sincrotrone Trieste, 34149 Basovizza, Italy.}}
\footnotetext{\textit{$^{6}$~Laboratoire de Chimie Physique--Matière et Rayonnement (LCPMR), UMR~7614, CNRS, Sorbonne Université, 75005~Paris, France.}}



\section{Introduction}
Water plays a fundamental role in a wide range of physical, chemical, and biological processes, making it one of the most important substances studied in science. Its relaxation dynamics under ionizing radiation is of particular importance to fields such as atmospheric chemistry,~\cite{falcinelli_possible_2015} astrochemistry,~\cite{van_interstellar_2013} radiation biology\cite{gao2025damaging} and radiation protection of ionized biomolecular systems.~\cite{johny2024water}

As isolated molecules, water exhibits well-characterized photoionization dynamics.~\cite{dutuit:1985, banna:1986, truong:2009, hans:2015, roos:2018} However, in hydrogen-bonded networks like water clusters or liquid water, relaxation dynamics can become far more complex and involve a variety of other processes. For instance, ionization of one \(\mathrm{H_2O}\) molecule in water clusters initiates proton transfer to a neighboring molecule, triggering a structural rearrangement of the cluster. This often leads to the formation of cluster ions \(\mathrm{OH(H_2O})_{n-1}\mathrm{H}^+\), which can subsequently evolve into more stable protonated cluster ions \(\mathrm{(H_2O})_{n-1}\mathrm{H}^+\) due to the loss of a hydroxyl radical~\(\mathrm{OH}\) from the cluster as a neutral fragment.~\cite{mizuse_structural_2011, tachikawa_ionization_2004, schnorr2023direct} These protonated cluster ions \(\mathrm{(H_2O})_{n-1}\mathrm{H}^+\) are typically the dominant ionic products observed by mass spectrometry of free water clusters prepared under warm conditions. However, mass spectrometry studies of water clusters prepared in cold environments indicate that both \(\mathrm{(H_2O})_{n-1}\mathrm{H}^+\) and \(\mathrm{(H_2O)}_{n}^+\) can be formed. The formation of \(\mathrm{(H_2O)}_{n}^+\) is attributed to the suppression of the OH radical loss, leading to stable \(\mathrm{OH(H_2O})_{n-1}\mathrm{H}^+\) cluster ions.\cite{mizuse:2011, gardenier:2009, shiromaru:1989, shinohara:1986} Under cold conditions, the excess internal energy deposited in the water cluster upon ionization can efficiently be dissipated in the surrounding environment, leading instead to evaporation of the weakly bound cold species rather than to ejection of the \(\mathrm{OH}\) radical,~\cite{jongma:1998, shinohara:1986} thereby stabilization of the initially ionized water cluster ions. 

To form \(\mathrm{(H_2O)}_{n}^+\) in abundances comparable to those of \(\mathrm{(H_2O})_{n-1}\mathrm{H}^+\), it is essential to efficiently dissipate the excess energy deposited in the water clusters upon their ionization by effective cooling processes. One effective approach is to embed water clusters in weakly interacting rare-gas matrices, such as helium nanodroplets (HNDs). HNDs, in particular, provide an ultracold (\(\sim 0.37\)~K), chemically inert, and superfluid environment that minimally perturbs the embedded dopants.~\cite{boatwright_helium_2013, toennies_helium_2022, yang_helium_2013} Moreover, HNDs suppress fragmentation of molecular clusters upon ionization as compared to direct photoionization of free clusters,~\cite{De_fragmentation_2024} offering a unique platform to study the spectroscopy of intact dimer/cluster cations.~\cite{tiefenthaler:2020,Zunzunegui-Bru:2023,foitzik:2025,foitzik:2026,gupta2025origin,ltaief:2020Xe}

Quantum chemical investigations of small unprotonated water clusters~\cite{dob13,herrts15,busch23} containing up to 5 molecules in isolated form have confirmed that they are most stable in their proton-transferred conformations, as OH(H$_2$O)$_n$H$^+$ compounds. However, recent infrared spectroscopy experiments~\cite{iguchi:2023} have shown that in HNDs the dimer can also be stabilized in its alternative conformer (H$_2$O-H$_2$O)$^+$, in which the two molecules are so-called hemibonded to each other, {\em i.e.} without involving a hydrogen bond. Hemibonding in the ionized water dimer is due to electronic exchange effects and has been specifically investigated by Busch and Sotoudeh.~\cite{busch23} Isomer coexistence is also well known in protonated clusters, with the prominent cases of the Zundel and Eigen cations~\cite{kulig:2014,prakash:2023} that persist as limiting cases even in bulk water.~\cite{marx99} As far as unprotonated clusters are concerned, the coexistence of the proton-transferred and hemibonded forms is anticipated in clusters larger than the dimer, where an broad structural diversity is expected in the HNDs.

Mass spectrometry and electron spectroscopy of free gas-phase water clusters has been extensively studied before upon photon or electron impact ionization.~\cite{belau_2007, lengyel_extensive_2014, hartweg_size-resolved_2017, ren_electron_2024} These investigations have yielded detailed information on, for example, the formation of cluster cations, size-resolved anisotropy parameters,~\cite{buck_size_2014, hartweg_size-resolved_2017} electronic structure of free water clusters,~\cite{ren_electron_2024} and evidence of secondary ionization induced by intermolecular decay processes.~\cite{gartmann_electron_2018, jahnke_ultrafast_2010, ren_ultrafast_2023} Penning ionization electron spectroscopy (PIES) of gas phase-water molecules has also been studied using metastable He~\cite{arfa_experimental_1994, haug_experimental_1985, yee_electron_1976} and Ne~\cite{falcinelli_possible_2015, brunetti_stereodynamics_2013, Brunetti_penning_2012} atoms using crossed atomic beams. However, PIES of water clusters formed in an ultracold, weakly perturbing environment—such as HNDs—remain largely unexplored to date.
Our recent results obtained on mass spectrometry of Penning ionized water clusters embedded in HNDs reveal that water cluster ions of various sizes can efficiently form inside HNDs with minimal fragmentation and that their formation depends highly on the HNDs size.~\cite{De_fragmentation_2024} Here, we present the first measurements on electron-ion coincidence spectra of Penning ionized small water clusters formed in HNDs, and compare them with those measured for free water clusters following direct photoionization. Although the measured droplet-correlated electron-ion coincidence spectra of water clusters are significantly broadened compared to those measured for free water clusters—due to scattering of the emitted electrons or ions inside the HNDs—they still provide valuable insights into the electronic orbitals that are involved in the ionization of water clusters and into whether the resulting cluster ions undergo fragmentation or not.

Our experiments are supported by dedicated quantum chemical calculations of water clusters, in which we assess their electronic stability from their lowest excitation energy, as obtained using time-dependent density-functional theory. Here we compare neutral, protonated, and unprotonated forms. Focusing on the unprotonated case in particular, we confirm the likely coexistence of hemibonded and proton-transferred conformers in clusters larger than the dimer, and find evidence for yet another type of conformer for the pentamer, with a clear OH component but a proton-shared complex. Proton-transferred conformers generally exhibit a very low first excitation energy, below 1~eV, which correlates with their highly transient nature and strong propensity to eliminate the OH radical in gas phase experiments.

\section{Experimental method}

The experiments on water cluster doped HNDs were conducted at the Gas-Phase beamline of the Elettra Synchrotron radiation (SR) facility, Trieste, Italy and at the AMO beamline of SR ASTRID2 in Aarhus, Denmark. Detailed descriptions of the experimental setup are available in previous works.~\cite{buchta_charge_2013, De_fragmentation_2024, bastian_new_2022} Briefly, HNDs were generated by expanding highly pressurized (50~bar) He gas through a cryogenically cooled nozzle ($5~\mu$m) into a vacuum. The size of the droplets, which ranged from $5100$ (droplet radius $R\approx 3.8$~nm) to $20,000$ ($R\approx 6.0$~nm) He atoms,~\cite{toennies_helium_2022} was controlled by varying the nozzle temperature between 19~K and 13~K.

The HNDs beam then passed through a doping cell where the $\mathrm{H_2O}$ or $\mathrm{D_2O}$ gas was introduced at a water partial pressure {$p_\mathrm{H_2O}=1.7 \times 10^{-5}$ and $4.6 \times 10^{-5}$, leading to an estimated average number of doped $\mathrm{H_2O}$ molecules $\bar{N}_\mathrm{H_2O}\approx 1$ and 2, respectively}, calculated using the model described in Ref.~\citenum{De_fragmentation_2024}. A mechanical chopper, mounted between the first skimmer and the doping cells, was used to interrupt the HNDs beam, and thus to measure the background signal due to photoionization of residual gas and free He atoms. The droplet-correlated signal can thus be obtained by discriminating the background signal from the droplet signal. After that, the droplet beam was doped with water molecules, and then entered the interaction chamber after passing through a second skimmer, where it crossed the SR photon beam. 

The interaction chamber is equipped with a velocity map imaging (VMI) spectrometer with an average energy resolution $\Delta{E}/E \approx 7$\% and a time-of-flight (TOF) spectrometer for detecting in coincidence electrons and ions emitted from the excited/ionized water doped HNDs. To infer total and coincidence electron spectra, electrons velocity map images were first recorded, and then Abel-inverted using MEVELER inversion method.\cite{dick_inverting_2014} The photon energy was fixed at $h\nu= 21.6$~eV while measuring the electron spectra and varied in the range of 20~eV to 22.5~eV while measuring the $h\nu$-dependent ion yields.

The experiment on photoionization of free water clusters was conducted at the AMO beamline of SR ASTRID2. The details of the experimental setup are given in Ref.~\citenum{bastian_new_2022}. A water cluster source similar to the one described in Ref.\citenum{forstel_source_2015} was used to produce a beam of free water clusters by supersonic expansion. It mainly consists of a water reservoir heated up to 95$^{\circ}$C and 80~$\mu$m diameter-nozzle maintained at a temperature of 130$^{\circ}$C higher than the temperature of the water reservoir to avoid its clogging. To record electron spectra and ion mass spectra from photoionized free water clusters, the photon energy was fixed at $h\nu= 20.6$~eV.
\section{Results and discussion}
We present in Fig.~\ref{fig: Mass_spectra_Ion_yields} mass spectra recorded for water clusters formed in HNDs and for free water clusters. The mass spectrum shown in panel a) of Fig.~\ref{fig: Mass_spectra_Ion_yields} is recorded for free water clusters at a photon energy of $h\nu=20.6$~eV. It exhibits a series of ion peaks attributed to protonated water cluster ions, i.e. $\mathrm{(H_2O)}_{n}{\rm H}^+$, that are formed following a proton transfer and a subsequent loss of OH upon direct photoionization of free water clusters:
\begin{equation}
\label{eq:photoionization of water clusters}
\mathrm{H_2O}_n + \mathrm{h\nu}\rightarrow  \mathrm{H_2O}_{n-1}\mathrm{H^{+}}\cdots\mathrm{OH} + e^-\rightarrow \mathrm{H_2O}_{n-1}\mathrm{H^{+}} + \mathrm{OH} + e^-
\end{equation}

Panel b) of Fig.~\ref{fig: Mass_spectra_Ion_yields} shows the mass spectrum of water clusters formed inside HNDs recorded at $h\nu=21.6$~eV for a droplet size $R\approx 4.8$ nm and average number of doped $\mathrm{H_2O}$ molecules $\bar{N}_\mathrm{H_2O}\approx 1$. It clearly shows distinct peaks arising from both protonated $\mathrm{(H_2O)}_{n-1}{\rm H}^+$ and unprotonated $\mathrm{(H_2O)}_{n}^+$ water cluster ions. The $\mathrm{(H_2O)}_{n-1}{\rm H}^+$ ions appear to be significantly more abundant than $\mathrm{(H_2O)}_{n}^+$ ions in the mass spectrum, consistent with our previous findings.\cite{De_fragmentation_2024} One notable advantage of forming cationic water clusters inside HNDs is the active cooling provided by the He environment. This efficient cooling helps to dissipate the excess internal energy stored during vertical ionization by evaporating some He atoms, thereby suppressing fragmentation such as OH loss. As a result, in addition to the formation of $\mathrm{(H_2O)}_{n-1}{\rm H^+}$ ions, a fraction of water cluster ions formed inside the HNDs upon ionization can be stabilized, leading to the formation of $\mathrm{(H_2O)}_{n}^+$. This is in sharp contrast to the case of photoionization of free water clusters where such stabilization does not occur and only $\mathrm{(H_2O)}_{n-1}{\rm H^+}$ ions are formed as a consequence of proton transfer and OH loss. In the mass spectrum shown in Fig.~\ref{fig: Mass_spectra_Ion_yields} b), both $\mathrm{(H_2O)}_{n-1}{\rm H^+}$ and $\mathrm{(H_2O)}_{n}^+$ ionic clusters are clearly visible up to $n-1 = 5$ and $n=5$, respectively, with $\mathrm{(H_2O)}_{n-1}{\rm H^+}$ being the more abundant. The relative yields of these ionic clusters ($n =2$--5) are shown in Fig.~\ref{fig: Mass_spectra_Ion_yields} c) as a function of $h\nu$ in the range $h\nu = 20.5$--22.5~eV. In this range, HNDs can either be resonantly excited into the droplet 1s2p $^1$P or 1s2S $^1$S absorption band where the 1s2p $^1$P band has a particularly large absorption cross-section (approximately 25~Mbarn per He atom) as compared to the 1s2s $^1$S band and other He droplet higher-lying bands (e.g. 1s3p/1s4p bands).~\cite{buchta2013extreme} Following resonant photoexcitation into the 1s2p $^1$P band, an ultrafast relaxation into the droplet metastable $1s2s$ $^1$S state often occurs within a timescale less than $\sim 0.25$~ps.\cite{mudrich_ultrafast_2020, laforge_relaxation_2022} The energy stored ($\sim 20.6$~eV) in the formed metastable $1s2s$ $^1$S excited He$^*$ atoms can in turn be either released radiatively through photon emission (fluorescence)\cite{vonhaeften:2011} or non-radiatively by Penning ionization, recently referred to as interatomic/molecular Coulombic decay (ICD). \cite{cederbaum_giant_1997,Jahnke:2020,ben_ltaief_charge_2019,laforge_interatomic_2024}

Penning ionization can lead to indirect ionization of either a foreign atom/molecule doped in the HNDs~\cite{ben_ltaief_charge_2019} or another excited He* that can form following multiple excitations of the droplets, particularly when the droplet size becomes sufficiently large (droplet size $R>6$~nm), as demonstrated recently in Ref.~\citenum{ltaief2024:PRR}. It turns out to be highly efficient in producing ions and electrons either from HND surface-bound alkali dopants or complex molecular species embedded inside HNDs.\cite{buchta_charge_2013, Shcherbinin_penning_2018, Asmussen_pyrimidine_2023,sen_electron_2024,ben_ltaief_charge_2019,mandal_penning_2020, laforge_interatomic_2024} The pronounced ion signals seen at $h\nu=21.6$~eV in Fig.~\ref{fig: Mass_spectra_Ion_yields} c) thus indicate that efficient indirect ionization of the doped water clusters occurs via Penning ionization following 1s$\rightarrow$2p photoexcitation of the HNDs, leading to protonated and cationic water cluster ions via these two reactions: 
\begin{figure}[H]
\centering
\includegraphics[width=0.6\linewidth]{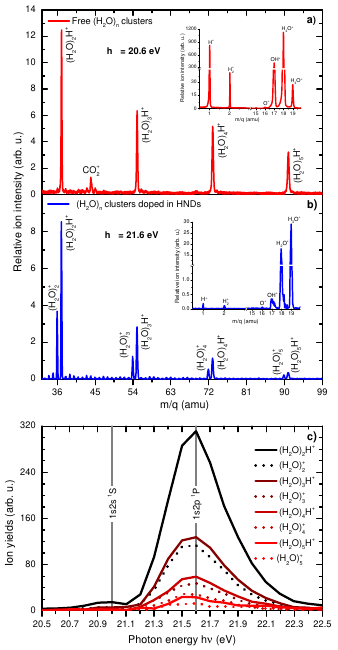}
\caption{\label{fig: Mass_spectra_Ion_yields} a) Ion mass spectrum recorded for free H$_2$O clusters at $h\nu$ = 20.6~eV. b) Ion mass spectrum of water clusters doped in HNDs measured at $h\nu$ = 21.6~eV for a droplet size $R$ $\approx$~4.8 nm and $\bar{N}_\mathrm{H_2O}$ $\approx$~1. The insets in top panel a) and b) show a close-up view of mass peaks corresponding to $\mathrm{H^+}$, $\mathrm{H_{2}^+}$, $\mathrm{O^+}$, $\mathrm{OH^+}$, $\mathrm{H_2O^+}$, $\mathrm{H_3O^+}$ ions. c) $h\nu$-dependent ion yields of $\mathrm{(H_2O)_{2-5}H^+}$ (solid lines) and $\mathrm{(H_2O)_{2-5}^+}$ (dotted lines) ions measured across the droplet 1s2s $^1$S and 1s2p $^1$P absorption bands in the exciting $h\nu$ range 20.5~eV--22.5~eV.} 
\end{figure}
\begin{equation}
\label{eq:H2On-ICD}
\begin{aligned}
\mathrm{He}^* + \mathrm{H_2O}_n\rightarrow  \mathrm{He} + \mathrm{H_2O}_{n-1}\mathrm{H^{+}}\cdots \mathrm{OH} + e^- \\
\rightarrow \mathrm{He} + \mathrm{H_2O}_{n-1}\mathrm{H^{+}} + \mathrm{OH} + e^-
\end{aligned}
\end{equation}
\begin{equation}
\label{eq:H2On-ICD}
\begin{aligned}
\mathrm{He}^* + \mathrm{H_2O}_n \rightarrow \mathrm{He} + \mathrm{H_2O}_{n-1}\mathrm{H^{+}}\cdots \mathrm{OH} + e^- \\
\rightarrow \mathrm{He} + \mathrm{H_2O}_{n}^{+} + e^-
\end{aligned}
\end{equation}

\begin{figure}[H]
\centering
\includegraphics[width=0.6\linewidth]{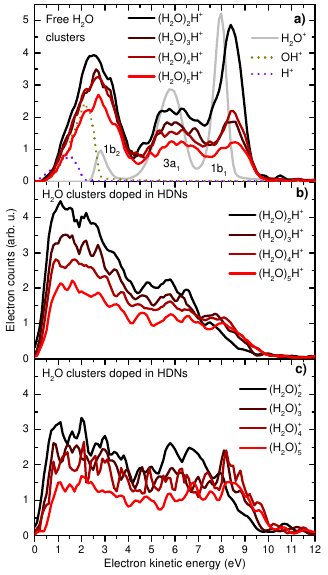}
\caption{\label{fig: Comparison_cluster_spectra_v4} a) Photoelectron spectra recorded in coincidence with $\mathrm{(H_2O)_{2-5}H^+}$ ions at $h\nu = 20.6$~eV for free water clusters. b) Droplet-correlated PIES's measured in coincidence with $\mathrm{(H_2O)_{2-5}H^+}$ at $h\nu = 21.6$~eV for water clusters doped HNDs ($R\approx 5.0$~nm). c) Droplet correlated PIES measured in coincidence with $\mathrm{(H_2O)_{2-5}^+}$ at $h\nu = 21.6$~eV for water clusters doped HNDs. The spectrum plotted in grey in panel a) shows a reference electron spectrum (scaled down by a factor of 100) measured for an effusive beam of isolated H$_2$O monomers at $h\nu = 20.6$~eV. The dark yellow and violet dashed lines (scaled down by a factor of 100) in panel a) show electron spectrum of OH$^+$ and H$^+$ ions created following direct photoionization of isolated water monomers, respectively.}
\end{figure}

The electronic configuration of an isolated H$_2$O molecule in its $^1\mathrm{A}_1$ ground state is: (1$\mathrm{a}_1)^2$(2$\mathrm{a}_1)^2$ 
(1$\mathrm{b}_2)^2$(3$\mathrm{a}_1)^2$(1$\mathrm{b}_1)^2$. 1$\mathrm{a}_1$ is a core orbital, primarly of O 1s character, and has a high binding energy of 539.8~eV.\cite{sankari2003vibrationally} 2$\mathrm{a}_1$, 1$\mathrm{b}_2$ and 3$\mathrm{a}_1$ are inner-valence orbitals and have binding energies of 32.4~eV, 18.7~eV and 14.8~eV,\cite{ning2008high, banna1986photoelectron, truong2009threshold} respectively. 1$\mathrm{b}_1$ is the outermost orbital and has an ionization potential of 12.6~eV.\cite{ning2008high, karlsson1975isotopic, truong2009threshold} The photon energy $h\nu = 20.6$~eV in reaction (1) or the energy ($\sim$ 20.6~eV) released from the metastable $1s2s$ $^1$S He$^*$ in reactions (2) and (3) is thus sufficient to ionize the water clusters and lead to ejection of an electron either from the 1b$_1$, 3a$_1$ or 1b$_2$ orbitals as their electron binding energies lie below 20.6~eV. The characteristic kinetic energy of an electron emitted via reactions (1), (2) and (3) is therefore the difference between 20.6~eV and the binding energy of an electron in the water clusters 1b$_1$, 3a$_1$ or 1b$_2$ orbital. The kinetic energy distributions of these emitted electrons measured in coincidence with their corresponding water cluster ions are shown in Fig.~\ref{fig: Comparison_cluster_spectra_v4}. Panels b) and c) show droplet-correlated PIES's recorded in coincidence with $\mathrm{(H_2O)_{2-5}H^+}$ and $\mathrm{(H_2O)_{2-5}^+}$, respectively, whereas the photoelectron spectra measured at $h\nu$=$20.6$~eV for free H$_2$O clusters are plotted in panel a) for a direct comparison. The gray spectrum shown in panel a) is a reference spectrum measured  at $h\nu$=20.6~eV for an effusive beam of gaseous H$_2$O molecules.
Upon direct photoionization of free-water clusters, emission of an electron out of the 1b$_1$, 3a$_1$ or 1b$_2$ orbitals occurs, resulting exclusively into the formation of protonated water cluster ions following reaction (\ref{eq:photoionization of water clusters}). Compared to the electron features of  an isolated H$_2$O, the free water cluster electron features appears a bit shifted in energy: the 1b$_1$ electron feature appears to be shifted in energy by $\sim 0.45$--0.65~eV [see Sec. 2 in supplementary material (SM)], whereas for the 3a$_1$ and 1b$_2$ electron features the shift in energy is less, i.e. $\sim$0.2~eV and $\sim$0.1~eV, respectively. These energy shifts are in good agreement with those reported in Ref.~\citenum{hartweg_size-resolved_2017} for free water clusters. Furthermore, the cluster-specific electron emission feature 1b$_2$ appears more pronounced and broader than the one seen in the electron spectrum of the H$_2$O monomer. This is mainly because the fragmentation channel that often leads to OH$^+$ and H$^+$ ions following 1b$_2$ photoionization of the free H$_2$O monomer is suppressed in water clusters. Note that the appearance energy of OH$^+$ and H$^+$ fragments following the removal of 1b$_2$ electron from a gaseous H$_2$O molecule is approximately 18.0~eV and 18.5~eV,\cite{roos:2018} respectively, which is about 2.5~eV and 2.1~eV lower than the photon energy $h\nu=20.6$~eV used in the present work to ionize the free water clusters.

\begin{figure}[h!]
\centering
\includegraphics[width=0.50\linewidth]{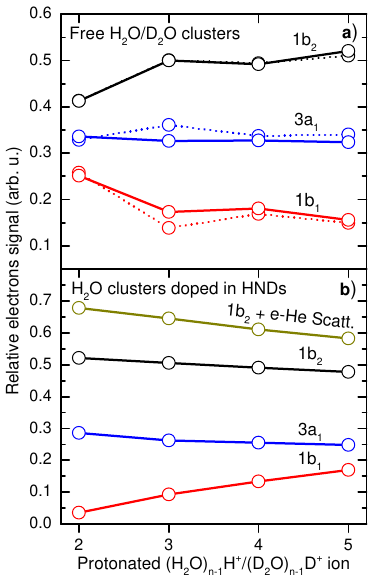}
\caption{\label{fig: Cross_section} a) Relative $1b_1$, $3a_1$, and $1b_2$ electron signals measured in coincidence with $\mathrm{(H_2O)}_{n-1}{\rm H^+}$ (solid lines with circles)/$\mathrm{(D_2O)_{n-1}D^+}$ (dotted lines with circles) for free H$_2$O/D$_2$O clusters, respectively, at $h\nu =20.6$~eV and as a function of $n=3$--6. b) Droplet correlated-relative $1b_1$, $3a_1$, and $1b_2$ electron signals measured in coincidence with $\mathrm{(H_2O)}_{n-1}{\rm H^+}$ ($n=3$--6) for H$_2$O clusters doped HNDs at $h\nu = 21.6$~eV and as a function of $n$. The solid gold line with circle in b) shows the sum of $1b_2$ electron signals (black solid line with circles) and the electron signal inferred from the generic e-He scattering spectrum. 
 } 
\end{figure}

For H$_2$O clusters doped in HND, the PIES's measured in coincidence with $\mathrm{(H_2O)_{2-5}H^+}$ [see Fig.~\ref{fig: Comparison_cluster_spectra_v4} b)] and $\mathrm{(H_2O)_{2-5}^+}$ [see Fig.~\ref{fig: Comparison_cluster_spectra_v4} c)] appear both broad and less structured relative to the electron spectra of free water clusters. For these PIES's, the 1b$_1$ and 3a$_1$ features remain almost distinguishable, but the 1b$_2$ feature is obscured by a broad signal that peaks at kinetic energy of $\sim 1$~eV and exhibits a broad tail extending up to $\sim 4$~eV. This low-kinetic energy feature ($<4$~eV) originates predominantly from electron-He scattering and can be referred to as a generic HND background spectrum. It has also been observed previously in nearly all PIES of atomic and molecular dopants embedded in medium-to large-sized HNDs, e.g. HNDs doped with heavier rare gases (Xe, Kr),\cite{wang2008photoelectron} and complex molecules such as acenes,\cite{Shcherbinin_penning_2018} coronene,\cite{ltaief_photoelectron_2021}, pyridine~\cite{Asmussen_pyrimidine_2023} camphor,\cite{sen_electron_2024} 
acetylene\cite{mandal_penning_2020} and aniline.\cite{loginov_photoelectron_2005} Despite the presence of this feature, valuable insights can still be extracted from the PIES's shown in Fig.~\ref{fig: Comparison_cluster_spectra_v4} b) and c). One noteworthy observation is, for example, that the 1b$_1$ electron feature in the PIES measured in coincidence with both $\mathrm{(H_2O)}_{n-1}{\rm H}^+$ and $\mathrm{(H_2O)}_{n}^+$ emerges for $n-1 = 2$ and $n= 2$, respectively, and becomes more and more visible as $n$ increases. Interestingly, the trend of this 1b$_1$ electron feature contrasts with that of the electron signal originating from the free water cluster 1b$_1$ orbital and which appears more pronounced for $n-1 = 2$ than for $n-1  > 2$, as shown in Fig.~\ref{fig: Comparison_cluster_spectra_v4} a). More details on these trends can be obtained from Figs.~\ref{fig: Cross_section} a) and b) that show the relative 1b$_1$, 3a$_1$ and 1b$_2$ electron signals of free water clusters and water clusters-doped HNDs as function of $n$, respectively. Fig.~\ref{fig: Cross_section} a) also presents relative 1b$_1$, 3a$_1$ and 1b$_2$ electron signals measured at $h\nu=20.6$~eV for free D$_2$O clusters, and which are inferred from the electron spectra shown in Fig. S1 of the SM.

All the relative signals shown in Fig.~\ref{fig: Cross_section} a) and b) are obtained by applying a multi-Gaussian fitting approach (see SM Fig. S5) to the electron spectra shown in Fig.~\ref{fig: Comparison_cluster_spectra_v4} a), Fig. S1 and Fig.~\ref{fig: Comparison_cluster_spectra_v4} b), and then normalizing them to the sum of the integrated 1b$_1$, 3a$_1$ and 1b$_2$ electron signals. The relative 1b$_1$, 3a$_1$ and 1b$_2$ electron signals obtained for free $\mathrm{H_2O}$ and $\mathrm{D_2O}$ clusters show almost the same trend. When $n-1$ increases from $n-1 = 2$ to $n-1 = 5$, the 1b$_1$ electron signal decreases, while the signal of electrons emitted out of 1b$_2$ increases and that of electrons emitted from the 3a$_1$ orbital remains almost constant. Interestingly, the opposite behavior is observed for $\mathrm{H_2O}$ clusters doped in HNDs, with the relative signal of electrons emitted from the 1b$_1$ orbital clearly increasing with increasing $n$ and that of electrons emitted from the 1b$_2$ orbital gradually decreasing when $n$ increases. The droplet correlated-relative 3a$_1$ electron signal shows only a slight decrease when $n$ increases. The dark yellow solid line with open circles shows the sum of the relative signal obtained from the generic HND spectrum with that of electrons ejected out of the 1b$_2$ orbital, which also decreases with increasing $n$.
The observed trend in the 1b$_1$ electron signal for free $\mathrm{H_2O}$ and $\mathrm{D_2O}$ clusters is attributed to fragmentation of cluster ions created following photoionization of initially large $\mathrm{H_2O}$/$\mathrm{D_2O}$ clusters. This in turn leads to ionic fragments of varying sizes, with $\mathrm{(H_2O)}_2{\rm H^+}$/$\mathrm{(D_2O)}_2{\rm D^+}$ being the most abundant species. Given the inherently broad size distribution of the water clusters being produced in free water cluster jet, one would expect that water clusters of various sizes are likely subjected to direct photoionization when the cluster jet is intersected by the synchrotron radiation beam, thereby leading to water cluster ions of different sizes. However, the electron spectra presented in Fig.~\ref{fig: Comparison_cluster_spectra_v4} a) do not show a clear cluster dependent-energy shift with increasing $n$. This suggests that the initial photoionization event predominantly involves a single motif of water clusters within the free water cluster beam, which in turn subsequently fragment into ionic cluster ions of varying sizes. Note that this fragmentation mechanism depends on the initial experimental conditions under which the water clusters are formed and can therefore lead to mass selected photoelectron spectra different from those presented in Fig.~\ref{fig: Comparison_cluster_spectra_v4} a), as previously observed in Ref.~\citenum{hartweg2019electron}. 

In contrast, the increase in the 1b$_1$ electron signal seen for $\mathrm{H_2O}$ clusters embedded in HNDs can be interpreted in terms of the suppression of this fragmentation mechanism. Upon Penning ionization, these clusters are ionized by emission of electrons from their 1b$_1$ orbitals; however, unlike free water clusters, the created cluster ions largely remain intact. This suppression of cluster ions fragmentation is mainly due to the highly efficient cooling provided by the surrounding helium environment, which rapidly dissipates excess energy through evaporation of surrounding He atoms instead of leading to cluster fragmentation. The trend of the 1b$_1$ electron signal presented in Fig.~\ref{fig: Cross_section} b ) thus reflects the enhanced detection of the emitted 1b$_1$ electrons in coincidence with intact cluster ions rather than with fragment ion cluster products. 

The overall signal of electrons ejected out of the 1b$_1$ orbital appear, however, weaker than that of electrons emitted out of the 3a$_1$ and 1b$_2$ orbitals, as it can be seen from Fig.~\ref{fig: Comparison_cluster_spectra_v4} b). This can be understood from the low fraction of the detected cluster ions produced upon the removal of 1b$_1$ electrons, as most of these cluster ions tend to reside inside the droplets rather than get ejected from the droplet surface. In contrast, Penning ionization of water clusters from their deeper-valence 3a$_1$ and 1b$_2$ orbitals creates electronically excited water cluster ions. Substantial excess energy becomes therefore internally available in the created water cluster ions upon the removal of these deeper-valence electrons, leading to their vibrational excitation. As a result, some of those vibrationally excited cluster ions become more mobile inside the droplets and can migrate more rapidly toward the droplet surface by a non thermal process in the course of their vibrational relaxation,\cite{smolarek:2010} making their subsequent ejection from the droplets and detection more favorable than those created upon the removal of 1b$_1$ electrons. The majority of these excited cluster ions likely relax upon leaving the droplet and may fragment in a manner similar to the cluster ions created from free water clusters, yielding smaller protonated cluster ions or a hydronium ion $\mathrm{H_3O}^{+}$, see SM Fig. S4.  

Furthermore, electrons emitted out of the 1b$_1$ orbitals have a kinetic energy of around 8~eV slightly higher than the lowest exitation energy ($\sim$7~eV~\cite{chipman:2005}) of individual water molecules. They can inelastically or elastically scatter with surrounding water molecules that form the water clusters inside the droplet, leading to their electronic and vibrational excitations.\cite{trajmar:1973, chutjian:1975, khakoo:2009, seng:1976} This can lead to a decrease in the signal of 1b$_1$ electrons peaking at around 8~eV, as well as an increase in the amount of scattered electrons of low kinetic energy, which can instead be detected in coincidence with the emitted cluster ions, thereby contributing to the broadening of the droplet-correlated PIES's presented in Fig.~\ref{fig: Comparison_cluster_spectra_v4} b) and c), see also SM Fig. S5. 

\begin{figure}[h!]
\centering
\includegraphics[width=0.6\linewidth]{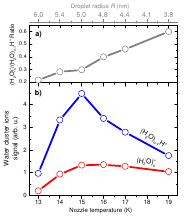}
\caption{\label{fig: ratio} a) Ratio of $(\mathrm{H_2O})_{2-5}^+$ to $(\mathrm{H_2O})_{2-5}\mathrm{H}^+$ cluster ions as a function of nozzle temperatures in the range of 13--19~K. b) Integrated $(\mathrm{H_2O})_{2-5}\mathrm{H}^+$ (blue line with open circle) versus $(\mathrm{H_2O})_{2-5}\mathrm{H}^+$ (red line with open circle) in the same nozzle temperature range. The scale on top of the upper panel indicates the initial mean HNDs radius $R$ determined according to Ref.~\citenum{toennies:2004}. 
} 
\end{figure}

\begin{figure}[h]
\centering
\includegraphics[width=0.8\linewidth]{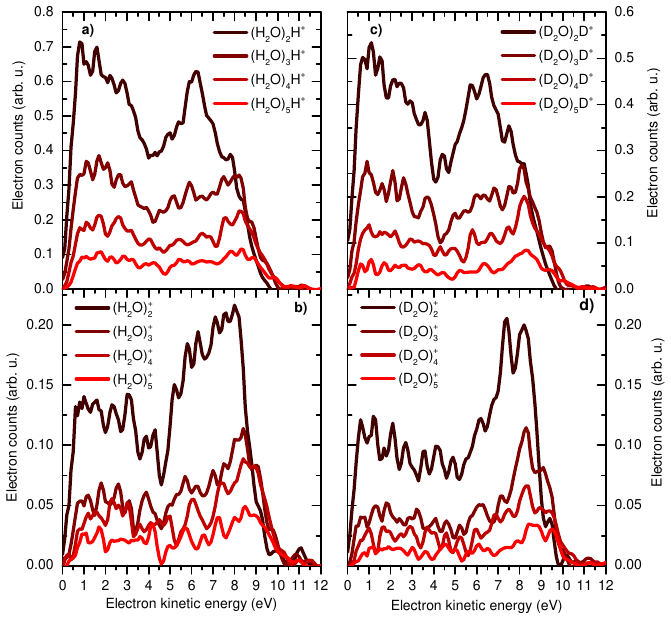}
\caption{\label{fig: HND_H2O_D2O_PIES} Droplet-correlated PIES's measured in coincidence with $\mathrm{(H_2O)_{2-5}H^+}$ [panel a)] and $\mathrm{(H_2O)_{2-5}^+}$ [panel b)] at h$\nu = 21.6$~eV for $\mathrm{(H_2O)_{n}}$ clusters doped small HNDs ($R$ $\approx$ 3.8~nm). Panels c) and d) show droplet-correlated PIES's measured in coincidence with $\mathrm{(D_2O)_{2-5}D^+}$ and $\mathrm{(D_2O)_{2-5}^+}$ at h$\nu = 21.6$~eV for $\mathrm{(D_2O)_{n}}$ clusters doped small HNDs. The mean number of $\mathrm{H_2O}$ and $\mathrm{D_2O}$ molecules doped in HND is estimated to be around 2.
} 
\end{figure} 

Fig.~\ref{fig: ratio}~a) shows the droplet size-dependent-ratio of the yield of $(\mathrm{H_2O})_n^+$ to $(\mathrm{H_2O})_{n-1}\mathrm{H}^+$ ions, recorded at $h\nu=21.6$~eV and for estimated average number of doped $\mathrm{H_2O}$ molecules $\bar{N}_\mathrm{H_2O}\approx~1$. As the droplet size decreases, this ratio increases. This can be understood from Fig.~\ref{fig: ratio}~b) which gives a direct comparison between the signals of $(\mathrm{H_2O})_{2-5}^+$ and $(\mathrm{H_2O})_{2-5}\mathrm{H}^+$ as a function of the droplet size. The $(\mathrm{H_2O})_{2-5}\mathrm{H}^+$ signal appears to be generally higher than the signal of $(\mathrm{H_2O})_{2-5}^+$ throughout the droplet size range. It increases from small droplet sizes ($R\approx 3.8$~nm), reaches a maximum at a medium droplet size ($R\approx 5$~nm), and then falls once the droplets become sufficiently large ($R > 5.4$~nm). The increase of the $(\mathrm{H_2O})_{2-5}\mathrm{H}^+$ signal is mainly due to an efficient proton transfer in the doped H$_2$O clusters as their size increases when the droplets become sufficiently large and efficient in picking up of many H$_2$O molecules.
However, the signal of $(\mathrm{H_2O})_{2-5}^+$ shows only a slight increase as the droplet size increases from $R\approx 3.8$~nm to $R\approx 5$~nm and then gradually decreases for larger droplets when the droplet size  is varied from $R\approx 5$~nm to $R\approx 6$~nm. This indicates that smaller HNDs where proton transfer becomes less efficient are particularly favorable for the detection of $(\mathrm{H_2O})_{n}^+$ ions.
The drop in the signal seen for both $(\mathrm{H_2O})_{2-5}\mathrm{H}^+$ and $(\mathrm{H_2O})_{2-5}^+$ in the droplet size range $R > 5$~nm is mainly due to the fact that most of those ions remain trapped inside the droplet and do not reach the droplet surface when the droplets become sufficiently large, thereby hindering their detection. This influence of droplet size on the likely position of the water clusters inside the droplet is supported by dedicated simulations of (neutral) water clusters using path-integral molecular dynamics simulations~\cite{christensen:2025}, described in the SM. These calculations show that clusters diffuse rather slowly inside the droplets, except when they initially lie in its outer parts, in which case their submersion takes place within about 100~ps but is significantly slowed down at subsequent times (see SM Sec. 7 and SM Fig. S8).

To what extent does the shape of the electron spectra of water clusters embedded inside HNDs change when the proton transfer contribution is minimal? To answer this question, we measured electron spectra of small HNDs (droplet radius $R=3.8$~nm) doped with H$_2$O/D$_2$O clusters at $h\nu=21.6$~eV and for an estimated average number of doped $\mathrm{H_2O}$ molecules of $\bar{N}_\mathrm{H_2O}\approx 2$, as shown in Fig.~\ref{fig: HND_H2O_D2O_PIES}. Panels a) and b) of Fig.~\ref{fig: HND_H2O_D2O_PIES} display PIES of H$_2$O clusters recorded in coincidence with $(\mathrm{H_2O})_{n-1}\mathrm{H}^+$ and $(\mathrm{H_2O})_n^+$, respectively. Panels c) and d) of Fig.~\ref{fig: HND_H2O_D2O_PIES} present PIES of D$_2$O clusters recorded in coincidence with $(\mathrm{D_2O})_{n-1}\mathrm{D}^+$ and $(\mathrm{D_2O})_n^+$, respectively. 
 
Both electron spectra recorded in coincidence with $(\mathrm{H_2O})_{n-1}\mathrm{H}^+$ and $(\mathrm{D_2O})_{n-1}\mathrm{D}^+$ exhibit similar structure; but their shapes are different than those presented in Fig.~\ref{fig: Comparison_cluster_spectra_v4} b) for large HNDs. Furthermore, their overall intensity gradually drops as $n$ increases. Moreover, the 3a$_1$ electron feature appears more pronounced and visible in the electron spectra measured in coincidence with $(\mathrm{H_2O})_2\mathrm{H}^+$/$(\mathrm{D_2O})_2\mathrm{D}^+$ than that seen for large HNDs [see blue spectrum shown in panel a) of Fig.~\ref{fig: Comparison_cluster_spectra_v4}]. However, for $n$ > 2, this 3a$_1$ electron feature becomes less visible, and the 1b$_1$ electron feature becomes the dominant one instead. 
 
Similarly the electron spectra measured in coincidence with $(\mathrm{H_2O})_n^+$ [see Fig.~\ref{fig: HND_H2O_D2O_PIES} b)] and $(\mathrm{D_2O})_n^+$ [see Fig.~\ref{fig: HND_H2O_D2O_PIES} d)] are comparable to one another but their common shape differs substantially from those presented in Fig.~\ref{fig: Comparison_cluster_spectra_v4} c) for large HNDs. Interestingly, the 1b$_1$ electron feature is almost the only clear and pronounced feature seen in all of these coincidence spectra, where its relative intensity appears much higher in the electron spectra measured in coincidence with  $(\mathrm{H_2O})_2^+$/$(\mathrm{D_2O})_2^+$ than in those measured in coincidence with $(\mathrm{H_2O})_{n>2}^+$/$(\mathrm{D_2O})_{n>2}^+$. The trend of this feature agrees well with the trend of the 1b$_1$ electron feature observed in the photoelectron spectra measured in coincidence with $(\mathrm{H_2O})_{2-5}\mathrm{H}^+$/$(\mathrm{D_2O})_{2-5}\mathrm{D}^+$ at $h\nu=20.6$~eV for free water clusters [see panel a) in Fig.~\ref{fig: Comparison_cluster_spectra_v4}]. However, it cannot be explained as due to fragmentation of larger Penning ionized water clusters into smaller ionic water cluster fragments, as interpreted above for the case of free water clusters, owing to the noticeable shift towards higher electron energies being found for the peak position of the 1b$_1$ feature seen in the electron spectra presented in Fig.~\ref{fig: HND_H2O_D2O_PIES} when $n$ increases from 2 to 5. The cluster size distribution of doped water clusters within HNDs depends strongly on the droplet size. This suggests that the pronounced 1b$_1$ electron feature seen in the PIES measured in coincidence with $(\mathrm{H_2O})_n^+$ originates instead from Penning ionized water clusters of various sizes formed inside the small HNDs, with $(\mathrm{H_2O})_2$ dimers being considerably more abundant than $(\mathrm{H_2O})_3$, $(\mathrm{H_2O})_4$, $(\mathrm{H_2O})_5$, etc. Penning ionization of water clusters doped in small HNDs thus leads to creation of more dimer ions over larger water cluster ions. Moreover, electron–He scattering is less efficient in smaller HNDs, as evidenced by the strong suppression of the generic low-energy feature seen in the spectra shown in Fig. 6, particularly for those recorded in coincidence with $(\mathrm{H_2O})_n^+$/$(\mathrm{D_2O})_n^+$. Electrons ejected by Penning ionization of water dimers/clusters doped small HNDs will thus undergo less scattering events, and their detection in coincidence with the corresponding emitted dimer/cluster water ions will further be enhanced. 
 
Note that the signature of direct photoionization of $(\mathrm{H_2O})_2$ leading to $(\mathrm{H_2O})_2^+$ has also been observed in earlier photoionization studies of free water clusters,\cite{Kamarchik_spectroscopic_2010, shiromaru:1987, hartweg_size-resolved_2017, ren_electron_2024} and only 1b$_1$ electron feature has also been seen in the electron spectrum recorded in coincidence with $(\mathrm{H_2O})_2^+$ ion.\cite{hartweg_size-resolved_2017, ren_electron_2024} It was interpreted as due to ionization of the hydrogen-bond donor water molecule that, together with another hydrogen-acceptor molecule, forms the water dimer. Calculations on the minimum energy structure of this free water dimer indicated that the valence orbital energies of the hydrogen donor and acceptor molecules are different; i.e. 11.92~eV and 13.45~eV, respectively.\cite{ren_electron_2024} A contribution from both hydrogen donor and hydrogen acceptor molecules may account for the considerable broadening of the 1b$_1$ electron peak seen in the PIES measured in coincidence with $(\mathrm{H_2O})_2^+$/$(\mathrm{D_2O})_2^+$, as shown in Fig.~\ref{fig: HND_H2O_D2O_PIES} b) and d). However, we cannot rule out the possibility that the water dimers formed inside the HND adopt the minimum energy structure reported for the free water dimer. This is because such a structure may either become stabilized or transform into another isomeric water dimer configuration inside HNDs, leading, for example, upon to its Penning ionization, to $(\mathrm{H_2O})_2^+$ with a $(\mathrm{H_3O})^{+}\cdot\mathrm{OH}$ or $\mathrm{H}^{+}\cdot\mathrm{(H_{2}O)}\cdot\mathrm{OH}$ cationic structure as a result of proton transfer accompanied with suppression of OH loss~\cite{mizuse:2011, mizuse:2013} or to the formation of the hemibonded water dimer cation $\mathrm{(H_{2}O\cdot\mathrm{OH_{2})^{+}}}$ as previously identified in Ref.~\citenum{iguchi:2023} under the HND environment. Furthermore, the 1b$_1$ electron peak seen in the PIES measured in coincidence with $(\mathrm{H_2O})_2^+$/$(\mathrm{D_2O})_2^+$ [Fig.~\ref{fig: HND_H2O_D2O_PIES} b) and d)] does not exhibit a clear energy shift relative to the 1b$_1$ peak position of the free water molecule [See SM Sec. 2] to clearly state that the water dimer is ionized at the hydrogen-donor or hydrogen-acceptor water molecule. It peaks around 8~eV and exhibits a broad tail toward the lower-kinetic energy part of the PIES extending up to 5 eV. The only clear energy shift seen for this droplet-correlated 1b$_1$ electron feature is that due to an increase in the doped cluster size when $n$ grows from $n=2$ up to $n = 5$.
 
In general, $(\mathrm{H_2O})_2^+$/$(\mathrm{D_2O})_2^+$ ions as well as $(\mathrm{H_2O})_2\mathrm{H}^+$/$(\mathrm{D_2O})_2\mathrm{D}^+$ ions and larger protonated or unprotonated water cluster ions are likely produced in large amounts via Penning ionization in both large or small HNDs. However, only a fraction of these cluster ions can be detected, as many of them tend to remain trapped inside the droplets. Upon their formation inside HNDs, water cluster ions may accommodate various structures as stable or electronically/vibrationally excited ionic species. Consequently, their detection efficiency strongly depends on their initial structure and electronic/vibrational excited states. Quantum chemistry  provides a valuable framework to get further insight into the possible structure of these water cluster ions, especially as they are formed from the neutral compounds, and on their relative electronic stability.

\section{Quantum chemical calculations}
\begin{figure*}[h!]
\centering
\includegraphics[width=16cm]{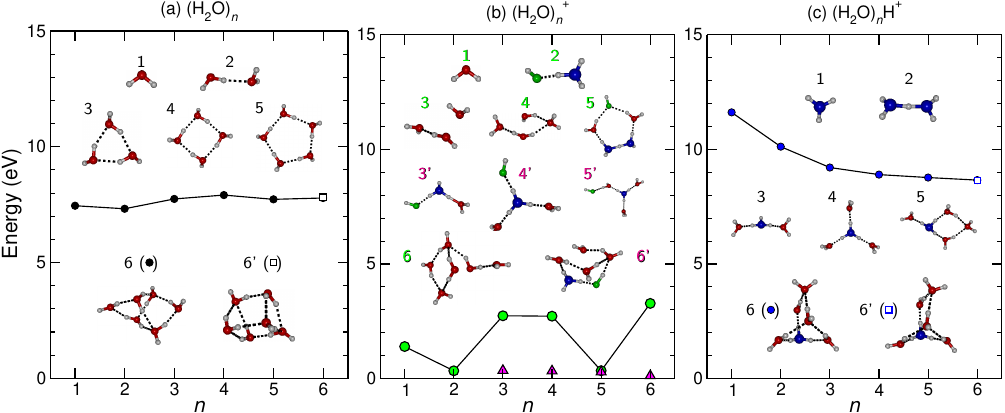}
\caption{\label{fig:QC} Calculated energies of the lowest electronic excited state of (a) neutral; (b) unprotonated; and (c) protonated water clusters as a function of their size $n=1$--6 (or the number of oxygen atoms), shown on a common scale. The corresponding structures are shown in each panel, with two isomers considered for the hexamers in the neutral and protonated cases, and for all clusters with $n>2$ for the unprotonated systems.}
\end{figure*}

A complete theoretical interpretation of the experimental ionization spectra inside the HNDs should imply a dynamical treatment of both nuclear and electronic degrees of freedom, ideally with a quantum account of the vibrations and the helium environment itself. Such an effort appearing unfeasible at this stage, we have reexamined one basic electronic feature of the water clusters likely to be formed in the experiment, namely their lowest electronic excitation energy associated with valence electrons. Focusing on unprotonated clusters, our goal is also to determine their likely structure under the cryogenic environment of the droplet, assuming they are initially in their stable neutral form, and to compare this structure to the known global minima,\cite{dob13,herrts15,busch23} which generally are of the proton-transferred form OH(H$_2$O)$_{n-1}$H$^+$.

The electronic structure of neutral, unprotonated, and protonated water clusters was thus computationally investigated, neglecting the helium environment, for sizes ranging from one to 6 oxygen atoms. More precisely, density-functional theory optimizations were conducted using the metahybrid M06-2X functional equipped with the aug-cc-pVDZ basis set, for the various following cases:
\begin{itemize}
    \item[(i)] Neutral water clusters (H$_2$O)$_n$, with their cyclic global minima for $n=3$--5 and the two competing cage and prism minima for the hexamer, with initial structures borrowed from the earlier quantum chemical exploration by Maheshwary and coworkers;\cite{maheshwary01}
    \item[(ii)] Unprotonated water clusters obtained upon direct ionization of these neutral structures;
    \item[(iii)] Alternative unprotonated clusters identified as putative global minima from the earlier work by Do and Besley,\cite{dob13} which all consist of distinct OH and H$_3$O subunits;
    \item[(iv)] Protonated water clusters with initial structures borrowed from the work by Hodges and Wales.\cite{hodgesw00} Here also a low-lying isomer was considered for the hexamer, in which one molecule flips one OH bond (isomer higher in energy by 110~meV).
\end{itemize}
All quantum chemical calculations were carried out using the Gaussian16 software package.\cite{g16} Once the various structures and isomers were successfully optimized with no imaginary frequency, their zero-point energy was evaluated in the harmonic approximation, without any scaling of the vibrational frequencies.

In the context of the present experiments, a common stability measure was employed to compare clusters across size, isomers and charge state, in the lowest excitation energy above the ground electronic state. For each system in its local or global minimum, single point time-dependent DFT calculations were thus carried out and the five lowest excitation energies recorded. The corresponding values are listed in Tables S1 and S2 of the SM.

Fig.~\ref{fig:QC} shows the variations of the lowest excitation energy for the different clusters, as a function of size and isomer. In the neutral clusters, the excitation energies only weakly oscillate in the 7--8 eV range and do not exhibit any significant variations upon isomerizing the hexamer. This suggests very homogeneous electronic structure among the different water molecules, a result that is confirmed by monitoring the charge distribution and its variance across the various oxygen atoms (see SM Fig. S6), as well as the ionization energy (see SM Fig. S7).

Protonated water clusters are not so homogeneous, because they usually carry one hydronium cation to form eigen complexes, except when $n=2$ where it is the Zundel cation which is preferred at the current DFT level, with a central proton equally shared by the two water molecules. However, the level of heterogeneity remains limited, and the resulting lowest excitation energy varies very smoothly with increasing cluster size, with almost no variation in the (very similar) secondary minimum of the hexamer. Interestingly, the lowest excitation energy converges slowly towards the same limiting value found in the neutral clusters, indicating the decreasing influence of the excess proton as more and more water molecules are incorporated.

In contrast, unprotonated clusters display not only generally low excitation energies, but also significant variations upon the isomer. Unprotonated water clusters directly obtained by removing one electron from the neutral yield structures that can be categorized either as (H$_2$O)$_n^+$ [cases of $n=3$, 4, and 6 in Fig. \ref{fig:QC}(b)] or as OH(H$_2$O)$_{n-1}$H$^+$ [cases of $n=2$ or 6' in Fig.\ref{fig:QC}(b)]. At the present level of DFT theory, the unprotonated dimer ion is of the proton-transferred form, whereas the trimer is hemibonded, with only one hydrogen bond between two molecules. The tetramer also misses one hydrogen bond, suggesting again hemibonding nature. The pentamer exhibits an even more complex structure with a distinct OH group, two water molecules, and a Zundel cation H$_5$O$_2^+$. A more complete characterization of this unprotonated cluster definitely appears desirable, using higher resolution approaches such as IR spectroscopy or anharmonic, wave-packet based, nuclear dynamics. From the perspective of their electronic stability, it is striking that the excitation energy of cationic water clusters strongly correlates with their chemical heterogeneity, with particularly low values ($<$1 eV) when a hydroxide group is present, and about 3~eV otherwise.

The above situation is relevant to clusters quenched upon being ionized, which is likely to happen in the cryogenic environment of the HNDs. These structures are significantly higher in energy than those identified by Do and Besley as the putative global minima,\cite{dob13} also shown in Fig.~\ref{fig:QC}(b) and labelled with primed numbers, all consisting of a proton transferred motif OH(H$_2$O)$_{n-1}$H$^+$. More precisely, the minima 3, 4, and 5 are found to be higher in energy than 3', 4', and 5' bt 1.02, 1.21, and 1.06~eV, respectively, including zero point energy in the harmonic approximation.
The lowest electronic excitation energies of these structures also differ markedly from those of the global minima, reflecting the key role of hydrogen delocalization. More importantly, the systematic presence of the OH group correlates with very low values of the excitation energy in the unprotonated clusters, which is consistent with their poor stability in mass spectrometry measurements under warm conditions, relative to protonated clusters. This confirms the need for a cold environment, such as the one provided by HNDs, to keep them alive in the experiment over long enough times.

\section{Conclusion}
In conclusion, we have studied the ionization mechanisms of small H$_2$O/D$_2$O water clusters doped in HNDs via Penning ionization and compared the results with those obtained for direct photoionization of free H$_2$O/D$_2$O water clusters. Upon Penning ionization, both (H$_2$O)$_{n-1}$H$^+$/(D$_2$O)$_{n-1}$H$^+$ and (H$_2$O)$_n^+$/(D$_2$O)$_n^+$ ions are observed in mass spectra. For larger HNDs, (H$_2$O)$_{n-1}$H$^+$ ions are found to be more dominant due to an efficient proton transfer process, whereas for smaller HNDs, proton transfer becomes less efficient and the detection of (H$_2$O)$_n^+$ ions is enhanced. In contrast, upon photoionization of free water clusters, only (H$_2$O)$_{n-1}$H$^+$ ions are observed, and the recorded coincidence spectra reveal three distinct electrons features corresponding to 1b$_1$, 3a$_1$ and 1b$_2$ bands, with \(1b_1\) electron feature appears to be most pronounced for $n= 2$ as a result of cluster ions fragmentation. For large HNDs, the electron signal associated with the orbital \(1b_1\) is strongly quenched in the PIES measured in coincidence with (H$_2$O)$_2$H$^+$/(H$_2$O)$_2^+$ as cluster ions fragmentation is suppressed inside the HNDs, whereas the \(3a_1\) and \(1b_2\) electron signals dominate the PIES's, although they appear broad due to scattering of the emitted electrons inside the HNDs with both surrounding neutral helium atoms and the water cluster aggregates. The \(1b_1\) electron feature becomes, however, more and more visible in the PIES's measured in coincidence with larger (H$_2$O)$_{n>3}$H$^+$/(H$_2$O)$_{n>2}^+$ as $n$ increases up to 5, indicating that intact water cluster ions of various sizes are directly created upon Penning ionization and not as a result of fragmentation of larger cluster ions. For small HNDs where proton transfer and electron-helium scattering are minimal, similar trends as for the case of large HNDs are also observed for \(1b_1\), \(3a_1\) and \(1b_2\) electron features seen in the PIES measured in coincidence with \(\mathrm{(H_2O})_{n-1}\mathrm{H}^+\)/\(\mathrm{(D_2O})_{n-1}\mathrm{D}^+\); however, in the PIES measured in coincidence with \((\text{H}_2\text{O})_{n}^+\)/\((\text{D}_2\text{O})_{n}^+\) only one clear \(1b_1\) electron feature is observed instead. This \(1b_1\) electron feature appears most pronounced in the electron spectrum measured in coincidence with $(\mathrm{H_2O})_2^+$/$(\mathrm{D_2O})_2^+$ and to shift in energy as the size of the doped water cluster increases. 

Quantum chemical calculations of the stable structures of protonated and unprotonated water clusters confirm the likely coexistence of proton-transferred and hemibonded structures. While the former are generally lower in energy, direct ionization of stable neutral clusters followed by cryogenic quenching in the helium environment should stabilize hemibonded cationic clusters not only in the dimer, but also potentially in the trimer and tetramer. More interestingly, our calculations predict that the pentamer cation should contain a proton-shared Zundel cation, and this opens interesting avenues for spectroscopic characterization of the kind recently pioneered by Iguchi and coworkers.\cite{iguchi:2023} Based on these quantum chemical calculations, time-resolved experiments could be performed in the future to efficiently probe the ejection mechanisms of stable water cluster ions formed inside HNDs. In such a pump–probe scheme, an extreme ultraviolet laser pulse would initiate the creation of the cluster ions by Penning ionization or direct photoionization of water clusters inside HNDs, while a delayed infrared or ultraviolet probe laser pulse would electronically excite them, thereby facilitating their ejection from the HNDs surface.

\section*{Author contributions}
{\textbf{Subhendu De}:} {conceptualization (equal); data curation (equal); formal analysis (equal); investigation (equal); methodology (equal); software (equal); visualization (equal); writing – original draft (equal); writing – review and editing (equal).}
{\textbf{Sivarama Krishnan}:} {conceptualization (equal); data curation (equal); investigation (equal); methodology (equal); writing – review and editing (equal).}
{\textbf{Niklas Sheel}:} {data curation (equal); formal analysis (equal); investigation (equal); methodology (equal); software (equal); visualization (equal); writing – review and editing (equal).}
{\textbf{Keshav Sishodia}:} {data curation (equal); formal analysis (equal); investigation (equal); methodology (equal); software (equal); visualization (equal); writing – review and editing (equal).}
{\textbf{Robert Richter}:} {data curation (equal); formal analysis (equal); investigation (equal); methodology (equal); resources (equal); software (equal); visualization (equal).}
{\textbf{Marcel Mudrich}:} {conceptualization (equal); data curation (equal); investigation (equal); methodology (equal); resources (equal); visualization (equal); writing – review and editing (equal).}
{\textbf{Florent Calvo}:} {data curation (equal); formal analysis (equal); investigation (equal); methodology (equal); software (equal); visualization (equal); writing – original draft (equal); writing – review and editing (equal).}
{\textbf{Ltaief Ben Ltaief}:} {conceptualization (equal); data curation (equal); formal analysis (equal); investigation (equal); methodology (equal); software (equal); visualization (equal);
project administration: (equal); supervision: (equal); writing – original draft (equal); writing – review and editing (equal)}

\section*{Conflicts of interest}
There are no conflicts to declare.


\section*{Acknowledgements}
L.B.L. acknowledges support by the Villum foundation via the Villum Experiment grant No. 58859. M.M. acknowledges support from the Novo Nordisk Foundation (grant no. NNF23OC0085401) and from Deutsche Forschungsgemeinschaft (DFG), grant no. 328961117 — SFB 1319 ELCH. S.R.K. and S. D. acknowledges support from the Dept. of Science and Technology, Govt. of India, the DST-DAAD scheme and Science and Eng. Research Board, and CEFIPRA (Indo-French Centre for the Promotion of Advanced Research). S.R.K. and S.D. acknowledge support from the Scheme for Promotion of Academic Research Collaboration, Min. of Edu., Govt. of India, and the Institute of Excellence programme at IIT-Madras via the Quantum Center for Diamond and Emergent Materials. S.R.K. acknowledge the Max Planck Society's Partner group programme.
The research leading to these results has been supported by the COST Action CA21101 ‘Confined Molecular Systems: From a New Generation of Materials to the Stars (COSY)’.

\balance

\providecommand*{\mcitethebibliography}{\thebibliography}
\csname @ifundefined\endcsname{endmcitethebibliography}
{\let\endmcitethebibliography\endthebibliography}{}

\end{document}


\maketitle
\vspace{-2cm} 
\noindent\large{
Subhendu De\textit{$^{1,6}$},
Sivarama Krishnan\textit{$^{1}$},
Niklas Sheel\textit{$^{2}$},
Keshav Sishodia\textit{$^{1}$},
Robert Richter\textit{$^{5}$},
Marcel Mudrich\textit{$^{2}$},
Florent Calvo\textit{$^{3}$},
Ltaief Ben Ltaief$^{\ast}$\textit{$^{2}$}
}

\vspace{0.3cm}

\noindent\textit{\small
$^{1}$ Quantum Center of Excellence for Diamond and Emergent Materials and Department of Physics, Indian Institute of Technology Madras, Chennai 600036, India.\\
$^{2}$ Department of Physics and Astronomy, Aarhus University, 8000 Aarhus C, Denmark.\\
$^{3}$ Université Grenoble Alpes, CNRS, LIPhy, 38000 Grenoble, France.\\
$^{5}$ Elettra-Sincrotrone Trieste, 34149 Basovizza, Italy.\\
$^{6}$ Laboratoire de Chimie Physique--Matière et Rayonnement (LCPMR), UMR~7614, CNRS, Sorbonne Université, 75005 Paris, France.
}

\section{Photoelectron spectra of free D$_2$O clusters}

\begin{figure}[h]
\centering
\includegraphics[width=0.6\linewidth]{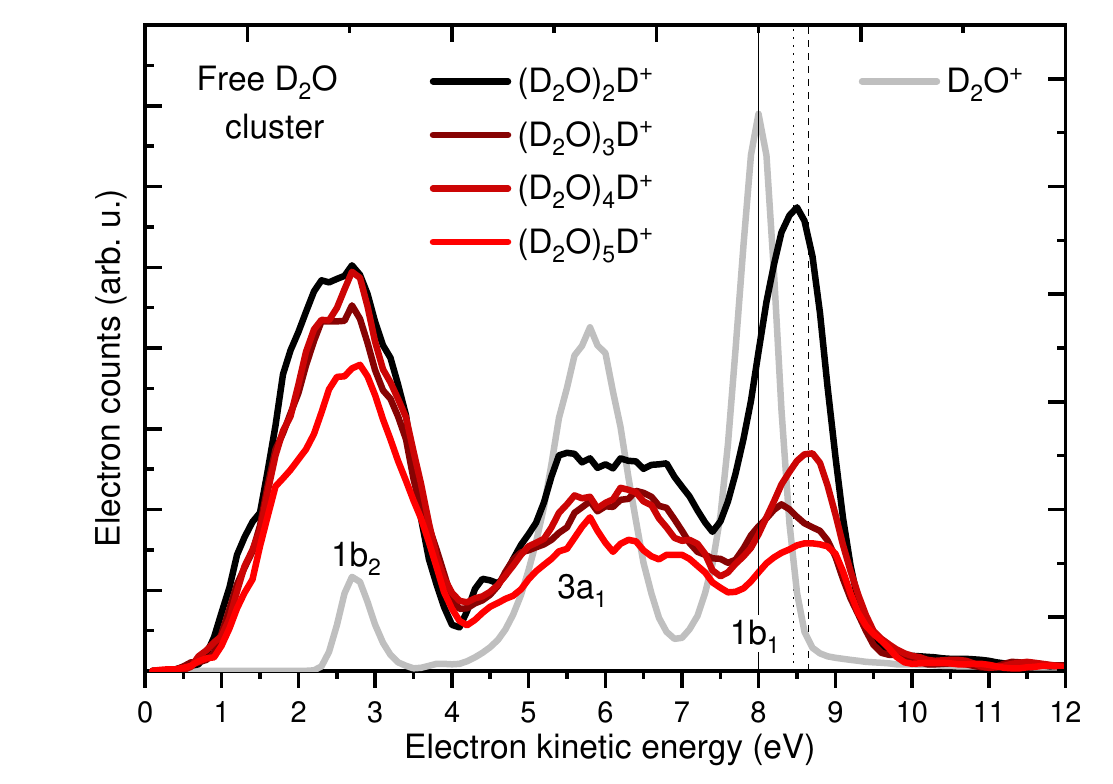}
\caption{\label{fig: Free D2O} Electron spectra measured in coincidence with $\mathrm{(D_2O)_{2-5}D^+}$ ions after photoionization of free $\mathrm{D_2O}$ clusters at a photon energy of 20.6~eV. The spectrum plotted in gray is a reference electron spectrum of free D$_2$O molecules recorded in coincidence with D$_2$O$^+$ at $h\nu = 20.6$~eV. The solid, dotted and dashed vertical lines indicate the energy positions of the 1b$_1$ peak seen in the electron spectra measured in coincidence with free $\mathrm{D_2O^+}$ molecular ion, $\mathrm{(D_2O)_{2}D^+}$ and $\mathrm{(D_2O)_{5}D^+}$, respectively.} 
\end{figure} 

The photoelectron spectra of free $\mathrm{D_2O}$ clusters measured at $h\nu = 20.6$~eV and in coincidence with $\mathrm{(D_2O)_{2-5}D^+}$ ions clearly show three peaks corresponding to 1b$_1$, 3a$_1$ and 1b$_2$ bands of $\mathrm{D_2O}$ cluster, see Fig.~\ref{fig: Free D2O}. The relative intensities of these different bands are discussed in the main text through Fig.~3 a) (dotted
lines with circles). The $\mathrm{1b_1}$ electron signal measured for free $\mathrm{D_2O}$ cluster is shifted by approximately $0.45$-$0.65$ eV (vertical lines in Fig.~\ref{fig: Free D2O}) relative to the $\mathrm{1b_1}$ electron signal recorded for free H$_2$O molecules (gray spectrum in Fig.~\ref{fig: Free D2O}).

\section{Shift in energy of the 1b$_1$ electron peak}

\begin{figure}[h!]
\centering
\includegraphics[width=0.43\linewidth]{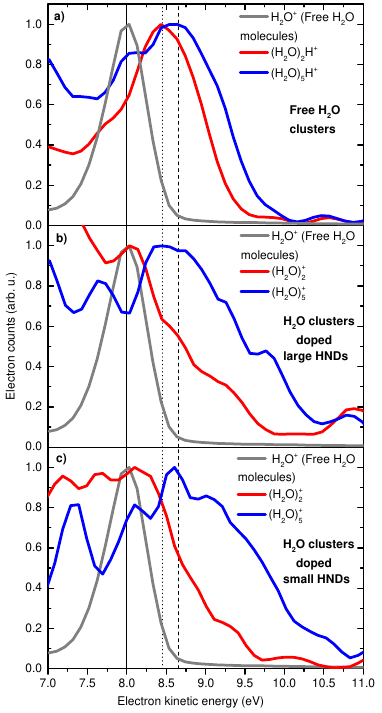}
\caption{\label{fig: shift_compare} Energy shift of the 1b$_1$ electron signal measured for the case of free H$_2$O clusters [a)] vs. H$_2$O clusters doped in large [b)] and small [c)] HNDs. The 1b$_1$ electron signals plotted in red and blue in a) are measured at $h\nu= 20.6$~eV in coincidence with (H$_2$O)$_2$H$^+$ and (H$_2$O)$_5$H$^+$, respectively. The droplet correlated-1b$_1$ electron signals plotted in red and blue in b-c) are measured at $h\nu= 21.6$~eV in coincidence with (H$_2$O)$_2^+$ and (H$_2$O)$_5^+$, respectively. The gray line is a reference electron spectrum of free H$_2$O molecules measured in coincidence with H$_2$O$^+$ at $h\nu = 20.6$~eV. The solid, dotted and dashed vertical lines indicate the energy positions of the 1b$_1$ electron peaks seen in the gray, red and blue spectra shown in a), respectively.}
\end{figure} 

In the electron spectra presented in Fig. 2 and Fig. 5 of the main text, the only noticeable energy shift is observed for the feature of the electrons emitted out of the 1b$_1$ orbital. Fig.~\ref{fig: shift_compare} compares the shift in energy of the $1b_1$ electron signal measured in coincidence with (H$_2$O)$_2^+$ and (H$_2$O)$_5^+$ for the case of H$_2$O clusters doped large [b)] and small [c)] HNDs with the corresponding energy shift of the $1b_1$ electron signal recorded in coincidence with (H$_2$O)$_2$H$^+$ and (H$_2$O)$_5$H$^+$ for free H$_2$O clusters [a)]. Relative to the $1b_1$ electron signal of free H$_2$O molecules [gray spectrum], the $1b_1$ feature measured for free H$_2$O clusters in coincidence with (H$_2$O)$_2$H$^+$ [red spectrum in a)] and (H$_2$O)$_5$H$^+$ [blue spectrum in a)] is shifted by approximately 0.45~eV and 0.65~eV, respectively. For H$_2$O clusters doped HNDs, however, the droplet-correlated $1b_1$ electron peak recorded in coincidence with (H$_2$O)$_2^+$ [red spectrum in b) and c)] is nearly unshifted relative to the $1b_1$ feature of free H$_2$O molecules [gray spectrum]. The clear shift in energy is observed only for the $1b_1$ electron broad feature seen in the electron spectrum recorded in coincidence with (H$_2$O)$_5^+$ [blue spectrum in b) and c)], which is of about 0.65~eV similar to that observed for free H$_2$O clusters.

\begin{figure}
\centering
\includegraphics[width=0.5\linewidth]{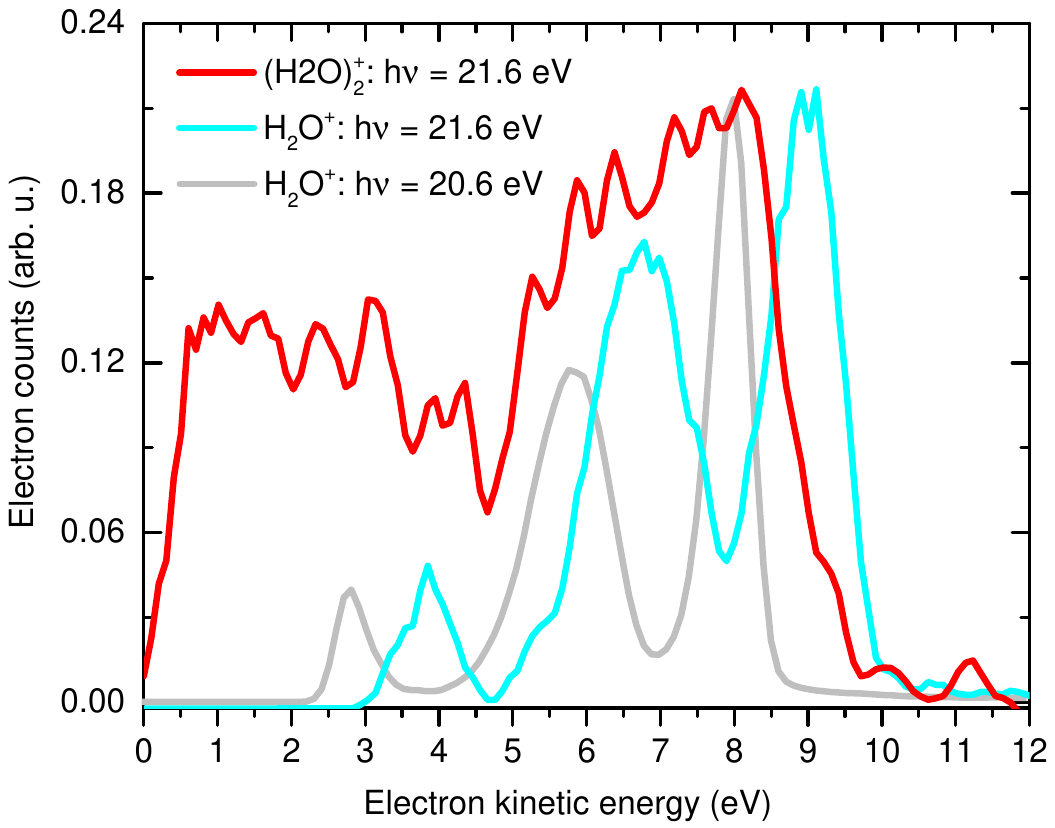}
\caption{\label{fig: compares_dimer} PIES (red line) measured in coincidence with (H$_2$O)$_2^+$ for H$_2$O clusters doped small HNDs at $h\nu = 21.6$~eV. The spectrum plotted in gray is a reference electron spectrum of free H$_2$O molecules measured in coincidence with H$_2$O$^+$ at $h\nu = 20.6$~eV. The spectrum plotted in cyan is an electron spectrum of undoped H$_2$O molecules measured in coincidence with H$_2$O$^+$ at $h\nu = 21.6$~eV while the HNDs beam is off.} 
\end{figure}

To check the energy calibration of all the droplet-correlated electron spectra presented in the main text and in the SM and those presented in Fig.~\ref{fig: shift_compare} to deduce the energy shift of the 1b$_1$ electron feature, reference electron spectra measured for undoped (droplet beam is off) H$_2$O/D$_2$O molecules in coincidence with H$_2$O$^+$/D$_2$O$^+$ are used, as shown in the example presented in Fig.~\ref{fig: compares_dimer}.




\section{Electron spectra in coincidence with protonated monomer ions}

\begin{figure}
\centering
\includegraphics[width=0.5\linewidth]{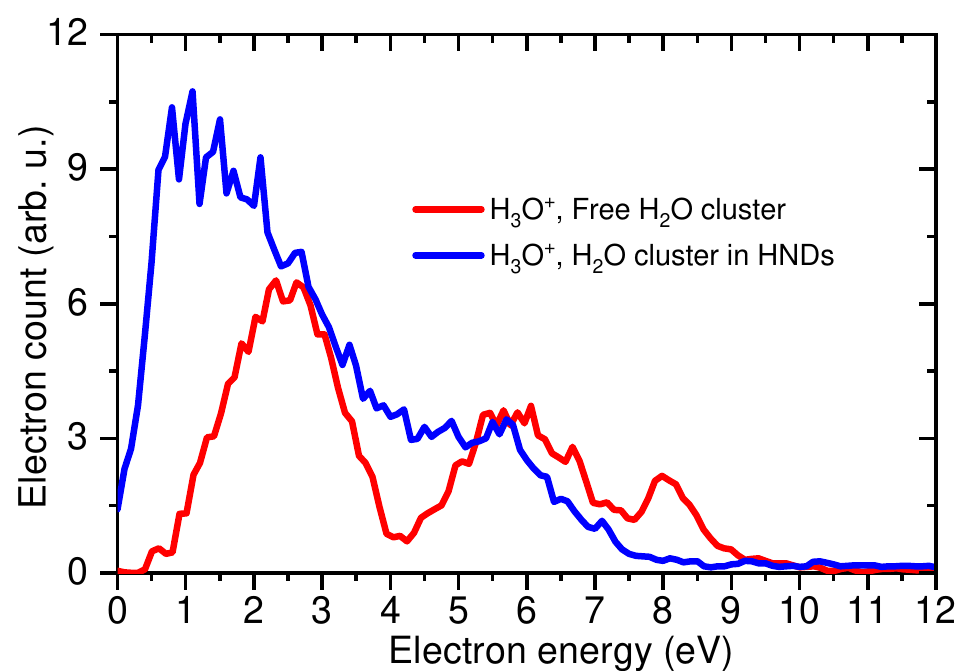}
\caption{\label{fig: compares_monomer} Electron spectra measured in coincidence with H$_3$O$^+$ at $h\nu = 21.6$~eV for Penning ionized water cluster doped HNDs (blue line) and at $h\nu = 20.6$~eV for photoionized free water clusters (red line). The blue spectrum is normalized accordingly to match the 3a$_1$ peak of the red spectrum.} 
\end{figure}

Fig.~\ref{fig: compares_monomer} compares the electron spectrum recorded in coincidence with H$_3$O$^+$ for free H$_2$O clusters at $h\nu = 20.6$~eV with that obtained from Penning ionization of HNDs doped with H$_2$O clusters. The photoelectron spectrum measured in coincidence with H$_3$O$^+$ following photoionization clearly exhibits three distinct bands corresponding to the 1b$_1$, 3a$_1$, and 1b$_2$ electron features. The 3a$_1$ and 1b$_2$ electron features are mainly due to fragmentation of electronically excited cluster ions upon direct photoionization of H$_2$O clusters. The 1b$_1$ electron characteristic that peaks at 8~eV appears unshifted in energy as compared to the 1b$_1$ electron peak of free H$_2$O molecules, see the grey spectrum shown in Fig. \ref{fig: shift_compare}. It thus reflects a contribution from the 1b$_1$ electron signal measured in coincidence with H$_2$O$^+$ after direct photoionization of single H$_2$O molecules that are present in the free water cluster jet as uncondensed molecules. The droplet-correlated Penning ionization electron spectrum (blue curve in Fig. \ref{fig: shift_compare}) does not exhibit this free H$_2$O monomer-1b$_1$ electron feature, and instead shows only two broad features; the 3a$_1$ and 1b$_2$ electron features. As discussed in the main text, the observation of only of these two features are attributed to Penning ionization from the inner-valence orbitals 3a$_1$ and 1b$_2$, which produces electronically excited cluster ions that are efficiently ejected from the HNDs and then undergo fragmentation, leading to either a hydronium H$_3$O$^+$ or to smaller water cluster ions. 

\section{Multi-Gaussian fitting approach}
\begin{figure}[h!]
\centering
\includegraphics[width=0.45\linewidth]{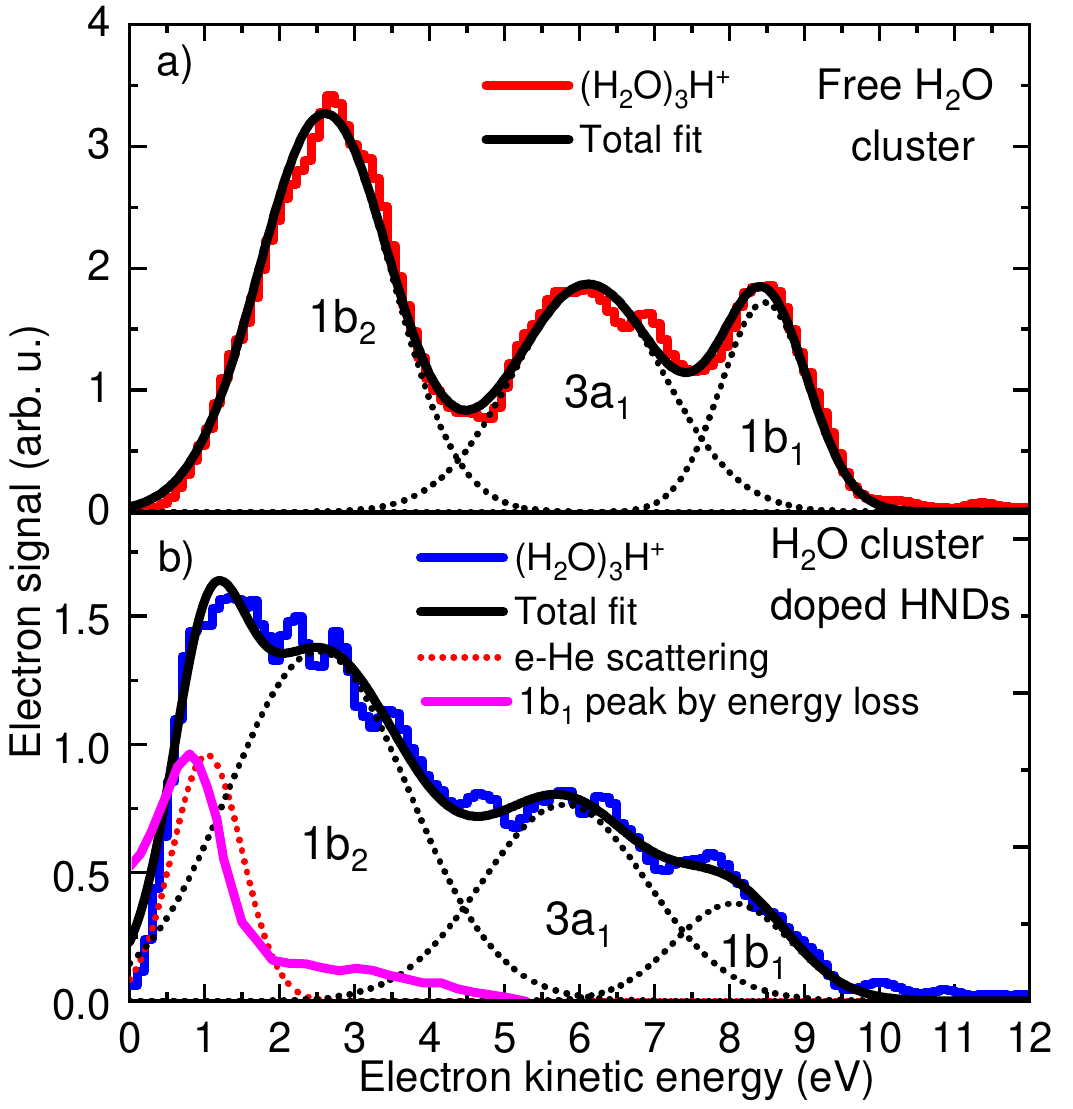}
\caption{\label{fig: Fit} a) Photoelectron spectra recorded in coincidence with (H$_2$O)$_3$H$^+$ for free water clusters at $h\nu= 20.6$~eV. b) PIES measured in coincidence with (H$_2$O)$_3$H$^+$ for water cluster doped HNDs ($R = 5$~nm) at $h\nu= 21.6$~eV.  Both electron spectra shown in a) and b) are fitted using a multi-Gaussian model: individual Gaussian fit for the 1b$_1$, 3a$_1$ and 1b$_2$ electron features are shown as dashed lines, while the resulting total fit is indicated by red and blue solid line in a) and b), respectively. The red dotted curve in b) is a Gaussian fit used to model the He-generic spectrum due to electron-He scattering. The pink solid curve in b) is electron impact excitation energy loss-spectrum of free H$_2$O molecules taken from~Ref. \citenum{trajmar:1973}.
} 
\end{figure}

Fig.~\ref{fig: Fit} shows an example of fitted electron spectra that are measured in coincidence with (H$_2$O)$_3$H$^+$ for both free water clusters [panel a)] and water clusters doped in HNDs [panel b)]. This figure mainly illustrates the three-Gaussian fitting approach used to extract information on the relative intensities of the 1b$_1$, 3a$_1$, and 1b$_2$ electron features seen in the electron spectra presented in the main text in Fig. 2 a) and Fig. \ref{fig: Free D2O} for free water cluster and in Fig. 2 b) for water cluster doped in HNDs. For water clusters doped in HNDs, an additional Gaussian fit [red dotted curve in b)] was also used to account for contribution from electron-He scattering.\cite{shcherbinin_penning_2018, Sen_camphor_2024} The integrated area of each fitted electron feature is normalized to the total area of the corresponding electron spectrum, yielding the relative intensities shown in Fig. 3 in the main text.

There is also a possibility that the H$_2$O molecules forming the doped water clusters inside HNDs can be electronically excited by the impact of the electrons emitted out of the 1b$_1$ orbitals by Penning ionization.\cite{trajmar:1973} The scattered electrons by these electron-impact excitation events may also contribute to the low-kinetic energy electron feature centered around 1~eV in the PIES of water cluster-doped HNDs. Their spectrum can take the shape of the pink solid line shown in b), which represents the energy loss spectrum of electrons after being inelastically scattered by free H$_2$O molecules.\cite{trajmar:1973} 
Note that such electron-impact excitation processes are expected to occur only in the presence of surrounding He atoms, where the initially outgoing electron can first undergo multiple scattering events with the He environment before colliding with neighboring H$_2$O molecules and then electronically exciting them. In contrast, for free water clusters, the outgoing electron can escape directly from the cluster, thereby strongly reducing the probability that such an electron impact excitation process will occur.



\section{Calculated excited state energies}

The electronic stability of neutral, unprotonated, and protonated water clusters containing up to 6 oxygen atoms was evaluated from the perspective of excited state energies. These energies were determined using time-dependent density-functional theory with the same functional (M06-2X) and basis set (aug-cc-pVDZ) as used to determine the ground state geometries, all calculations being performed also with the Gaussian16 software package.\cite{g16}

The list of 5 lowest excited state energies and the associated oscillator strengths are reported in Table \ref{tab:XSE1} for the clusters of the main series in Fig. 6 of the main article, and in Table \ref{tab:XSE2} for other isomers quoted with a prime in the same figure.

\begin{table}
\begin{tabular}{c||c|c||c|c||c|c}
  \hline \hline
  Number of & \multicolumn{2}{|c||}{(H$_2$O)$_n$} & \multicolumn{2}{|c||}{(H$_2$O)$_n^+$} & \multicolumn{2}{|c}{(H$_2$O)$_n$H$^+$} \\
  water molecules & $E_i$ (eV) & $f$ & $E_i$ (eV) & $f$ & $E_i$ (eV) & $f$ \\
  \hline
    & 7.4537 & 0.0427 & 1.3985 & 0.0003 & 11.6150 & 0.0400 \\
    & 8.9209 & 0.0000 & 5.6587 & 0.0000 & 13.7316 & 0.0105 \\
  1 & 9.5904 & 0.0795 & 12.4902 & 0.0102 & 13.7320 & 0.0105 \\
    & 10.3001 & 0.0019 & 13.4702 & 0.0168 & 16.4517 & 0.0000 \\
    & 11.0165 & 0.0115 & 13.7488 & 0.0662 & 16.4729 & 0.0001 \\
  \hline
    & 7.3219 & 0.0372 & 0.3398 & 0.0000 & 10.1185 & 0.0508 \\
    & 7.8728 & 0.0523 & 3.8617 & 0.0006 & 7.8334 & 0.0006 \\
  2  & 8.1532 & 0.0022 & 7.8334 & 0.0034 & 11.8255 & 0.1541 \\
    & 8.9861 & 0.0018 & 9.6289 & 0.0192 & 11.9246 & 0.0021 \\
    & 9.2275 & 0.0314 & 10.3682 & 0.0066 & 11.9832 & 0.0053 \\
  \hline
    & 7.7474 & 0.0397 & 2.7398 & 0.0102 & 9.2068 & 0.0142 \\
    & 7.8915 & 0.0683 & 4.0708 & 0.0794 & 9.2225 &0.0637 \\
  3  & 7.9325 & 0.0275 & 5.1480 & 0.1031 & 10.4625 & 0.0354 \\
    & 8.9910 & 0.0010 & 5.4047 & 0.0933 & 10.7901 & 0.0517 \\
    & 9.0377 & 0.0061 & 6.0331 & 0.0003 & 10.8313 & 0.0006 \\
  \hline
    & 7.9132 & 0.0743 & 2.7226 & 0.0067 & 8.8987 & 0.0069 \\
    & 7.9133 & 0.0743 & 4.2657 & 0.0808 & 9.9169 & 0.0555 \\
  4  & 7.9191 & 0.0440 & 4.6902 & 0.0648 & 8.9173 & 0.0558 \\
    & 7.9397 & 0.0000 & 5.1046 & 0.0825 & 10.5548 & 0.0083 \\
    & 8.8817 & 0.0000 & 5.4766 & 0.0127 & 10.5554 & 0.0082 \\
  \hline
    & 7.7302 & 0.0222 & 0.3450 & 0.0000 & 8.7704 & 0.0375 \\
    & 7.8931 & 0.0209 & 3.6588 & 0.0006 & 8.8061 & 0.0206 \\
  5  & 7.9080 & 0.0402 & 5.4650 & 0.0021 & 8.9018 & 0.0479 \\
    & 7.9214 & 0.0433 & 5.5118 & 0.0021 & 9.9539 & 0.0571 \\
    & 7.9389 & 0.0896 & 7.0564 & 0.0001 & 10.1967 & 0.0004 \\
  \hline
    & 7.7866 & 0.0332 & 3.2800 & 0.0128 & 8.6603 & 0.0201 \\
    & 7.9556 & 0.0839 & 3.7876 & 0.0007 & 8.7268 & 0.0440 \\
  6 & 8.0189 & 0.0073 & 4.7300 & 0.1719 & 8.8070 & 0.0099 \\
    & 8.0844 & 0.0336 & 4.8958 & 0.0234 & 8.8942 & 0.0815 \\
    & 8.1531 & 0.0965 & 5.2104 & 0.0011 & 8.9165 & 0.0573 \\
  \hline
  \hline
\end{tabular}
\caption{Energies of the five lowest excited electronic states, relative to the ground state, for neutral, unprotonated, and protonated water clusters of the main series in Fig. 6, as predicted by TDDFT calculations. The oscillator strengths $f$ are also provided.}
\label{tab:XSE1}
\end{table}

\begin{table}
\begin{tabular}{c||c|c||c|c||c|c}
  \hline \hline
  Number of & \multicolumn{2}{|c||}{(H$_2$O)$_n$} & \multicolumn{2}{|c||}{(H$_2$O)$_n^+$} & \multicolumn{2}{|c}{(H$_2$O)$_n$H$^+$} \\
  water molecules & $E_i$ (eV) & $f$ & $E_i$ (eV) & $f$ & $E_i$ (eV) & $f$ \\
  \hline
  \hline
    & & & 0.3528 & 0.0000 & & \\
    & & & 3.7472 & 0.0007 & & \\
  3'& & & 7.3418 & 0.0004 & & \\
    & & & 7.6509 & 0.0023 & & \\
    & & & 9.0111 & 0.0167 & & \\
  \hline
    & & & 0.3313 & 0.0000 & & \\
    & & & 3.7109 & 0.0008 & & \\
  4'& & & 6.9929 & 0.0000 & & \\
    & & & 7.0645 & 0.0008 & & \\
    & & & 7.3678 & 0.0018 & & \\
  \hline
    & & & 0.2881 & 0.0000 & & \\
    & & & 3.6785 & 0.0008 & & \\
  5'& & & 6.2770 & 0.0032 & & \\
    & & & 7.5468 & 0.0000 & & \\
    & & & 7.7955 & 0.0000 & & \\
  \hline
    & 7.8219 & 0.0555 & 0.1238 & 0.0000 & 8.6474 & 0.0318 \\
    & 7.9895 & 0.0323 & 3.1504 & 0.0005 & 8.7332 & 0.0389 \\
  6'& 8.0547 & 0.0081 & 5.3320 & 0.0396 & 8.8291 & 0.0170 \\
    & 8.1425 & 0.0275 & 5.9103 & 0.0002 & 8.8662 & 0.0672 \\
    & 8.1666 & 0.1424 & 5.9622 & 0.0011 & 8.9161 & 0.0554 \\
  \hline
  \hline
\end{tabular}
\caption{Energies of the five lowest excited electronic states, relative to the ground state, for the prism isomer of neutral, unprotonated, and protonated water clusters and the lowest-energy unprotonated minima with 3-5 water molecules, as predicted by TDDFT calculations. The oscillator strengths $f$ are also provided.}
\label{tab:XSE2}
\end{table}

\clearpage
\section{Electronic structure properties: charge inbalance and ionization potential}

Besides excited state energies, we have considered two alternative indicators of the electronic stability of water clusters in their neutral, unprotonated, and protonated form.

A global charge inbalance indicator was introduced as the variance between the partial charges carried by all oxygen atoms in the cluster. Partial charges themselves are derived from the ground state electronic structure obtained at the DFT level, using natural bond orbital partitioning.

\begin{figure}[h]
\centering
\includegraphics[width=0.7\linewidth]{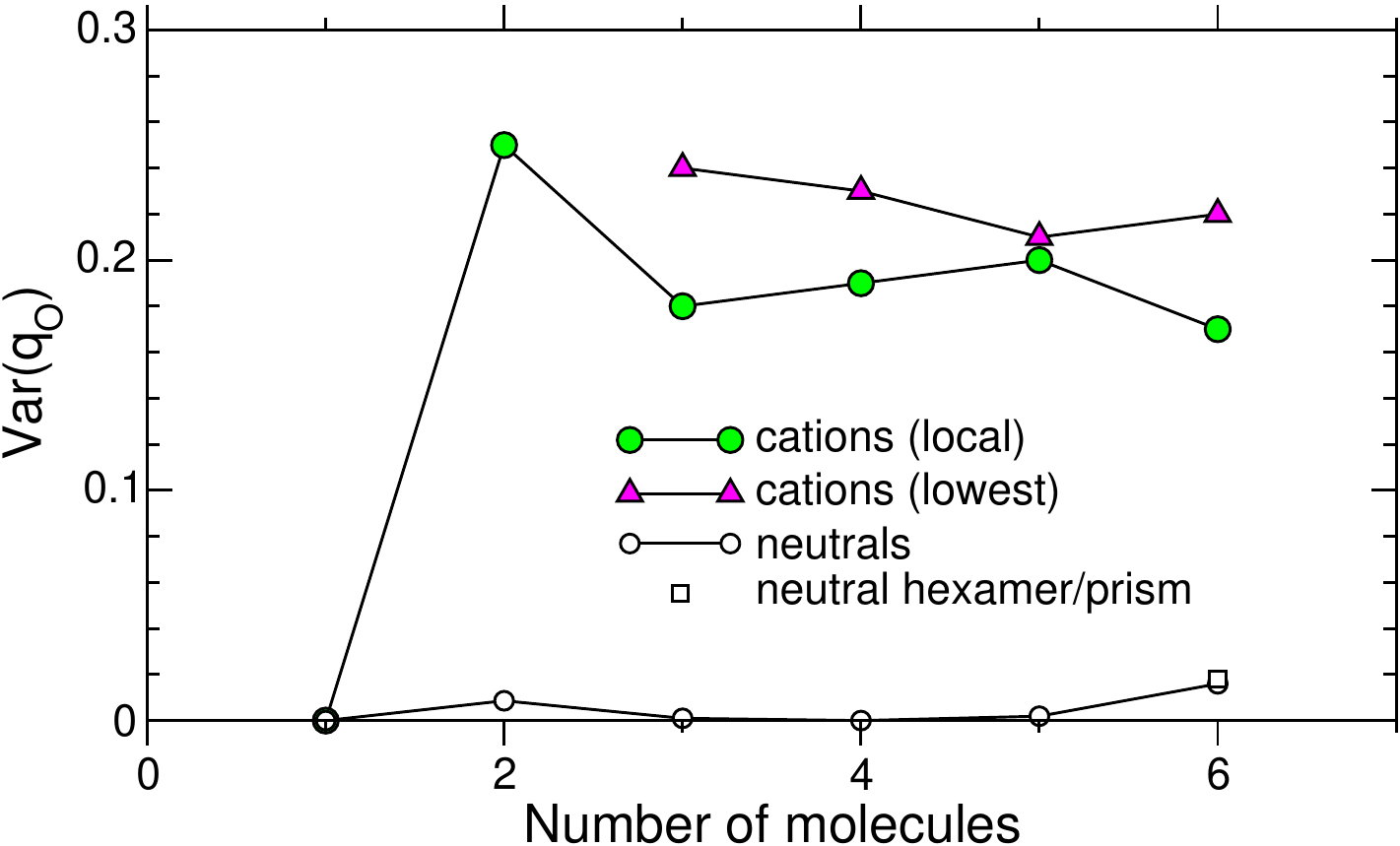}
\caption{Variance of the NBO charge on oxygen atoms in neutral water clusters (black circles) and in unprotonated clusters obtained by local minimization from the neutrals (green circles), compared with the lowest energy minima of the cations (yellow triangles). The black square shows the value for the neutral hexamer in the prism isomer. Charges are defined in units of the electron charge.} 
\label{fig:NBOcharge}
\end{figure}

The variations of this indicator with increasing cluster size are shown in Fig.~\ref{fig:NBOcharge} for the unprotonated clusters locally minimized from the neutral geometries, or in their lowest-energy configuration, and in comparison with the neutrals. 

While individual molecules are effectively neutral in the neutral cluster, with almost uniform charge distribution, a strong inbalance is found for all unprotonated clusters. This is not a surprise for clusters consisting of clear OH and H$_3$O units, for which effective charges closer to -e and +e, respectively, are found up to some overall dilution of the global +e charge. However, a comparable inbalance is obtained for all unprotonated clusters with 2 molecules or more, even when they do not contain clear OH and H$_3$O units as in the cases of $n=3$, 4, and 6 (cage) as obtained from local minimization of the corresponding neutral clusters. This result is consistent with the structural asymmetry of these latter clusters. While it does not particularly points at the specific molecules that cause this inbalance, this indicator confirms that water molecules form a rather heterogeneous set in the unprotonated clusters.

\begin{figure}[h]
\centering
\includegraphics[width=0.7\linewidth]{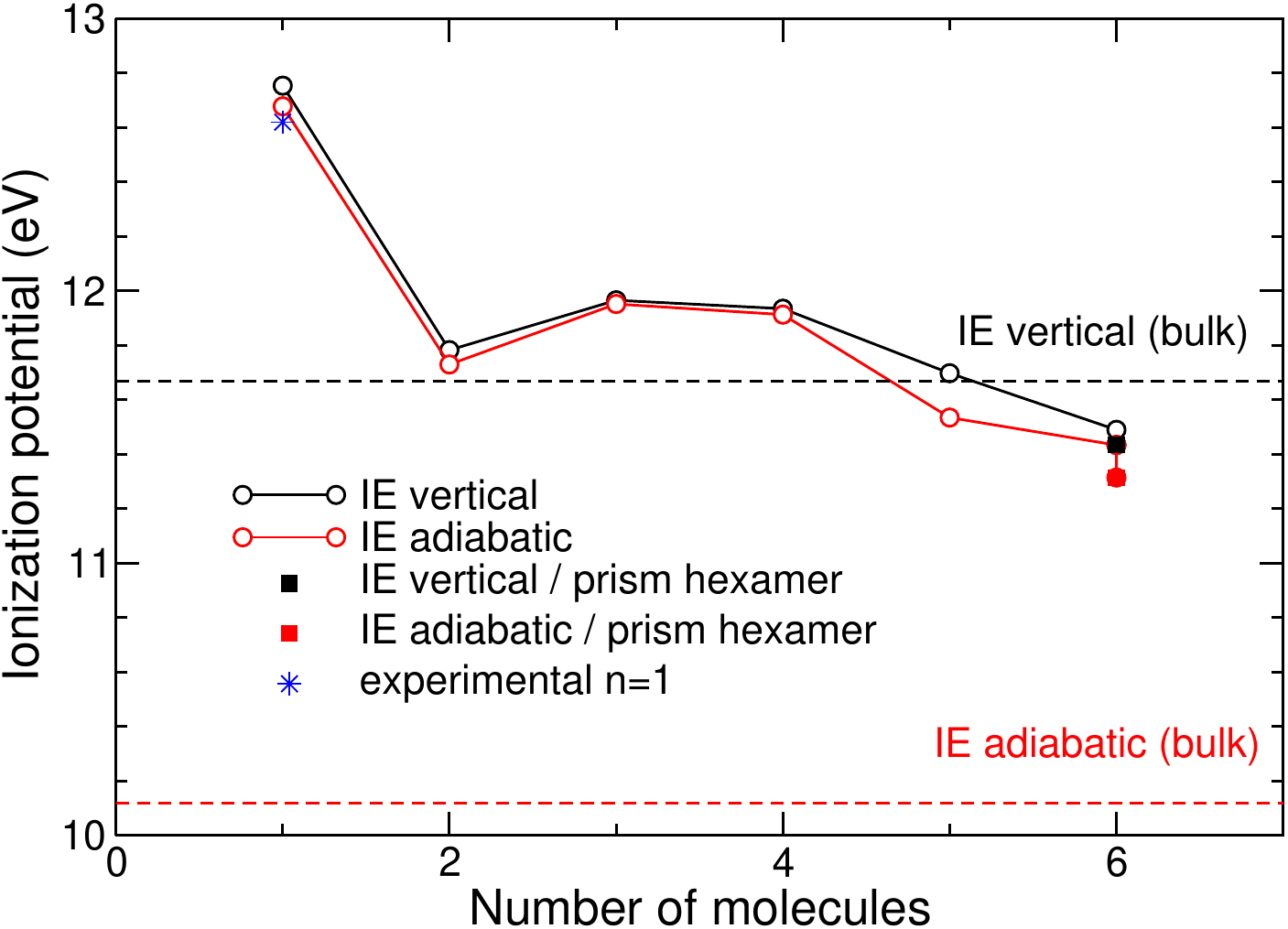}
\caption{Ionization energies of water clusters, as obtained from DFT quantum chemical calculations. The vertical and adiabatic values are represented as black and red circles, respectively, with full squares showing the values for the prism isomer of the hexamer. The experimental ionization energy for the water monomer is highlighted with a blue star, while the vertical and adiabatic ionization energies for bulk liquid water are indicated by dashed horizontal lines.} 
\label{fig:IE}
\end{figure}

Ionization energies have also been computationally determined using the same density-functional theory method, and Fig.~\ref{fig:IE} shows the variations of the vertical and adiabatic values across the size range of 1--6 water molecules. For the hexamer, both the cage and prism isomers were considered.
For the monomer, our results ($\sim$12.6--12.8 eV) reproduce rather well the known experimental value of 12.6 eV,\cite{snow90} setting the accuracy of the computational method. In clusters, both the vertical and adiabatic ionization potentials drop significantly and lie in the 11.5--12~eV range, with a minor decreasing trend but no major size effect. It is also interesting to compare with the bulk limit of liquid water,\cite{conaill20} for which the vertical (11.7~eV) and adiabatic (10.1~eV) ionization energies differ quite a lot. While the vertical values match well already for small clusters, the residual difference in the adiabatic case suggests major structural rearrangements in the condensed phase that are not captured by the gas phase clusters.

\section{Path-integral molecular dynamics simulations of the water tetramer in small helium nanodroplets}

Path-integral molecular dynamics (PIMD) simulations were conducted to simulate the diffusion process of neutral water clusters inside helium droplets. The TIP3P flexible potential was used to describe intra- and intermolecular bonding in the water cluster, and simple Lennard-Jones (LJ) potentials provided the appropriate framework to model the interactions among helium atoms, or between helium and both hydrogen and oxygen atoms. The distance parameters of these LJ interactions read $\sigma_{\rm He-He}=2.635$~\AA, $\sigma_{\rm He-O}=2.949$~\AA, and $\sigma_{\rm He-H}=2.620$~\AA, respectively, while the well depths are $\varepsilon_{\rm He-He}=0.93$~meV, $\varepsilon_{\rm He-H}=0.67$~meV, and $\varepsilon_{\rm He-O}=2.65$~meV. Note that, for the mixed He-H and He-O interactions, these values are obtained using the standard Lorentz-Berthelot combination rules. To better describe the electrostatic component of the interactions, a vectorial polarization potential was added to each helium atom, in response to the electric field created by the partial charges on the hydrogen and oxygen atoms of the water cluster. An isotropic polarizability parameter of 0.2047~\AA$^3$ (experimental value) was employed for helium.

The 1000-atom helium droplets were initially thermalized at 2~K using conventional PIMD trajectories, in the absence of the water cluster and using a modest Trotter discretization of $P=32$ that is sufficient to ensure a liquid (but not superfluid) medium. A time  step of 0.5~fs was used in the simulations, whose trajectories were propagated using massive Nosé-Hoover thermostatting, following details of that can be found in earlier publication.\cite{habershon13} This thermalization procedure was continued for 10~ps, after which the water tetramer was inserted at a specified distance from the droplet center of mass, moving helium atoms that were too close ($<$3~\AA) from the cluster to the outer border of the droplet. The PIMD simulation was then resumed and further propagated for one nanosecond.

\begin{figure}[h]
\centering
\includegraphics[width=0.7\linewidth]{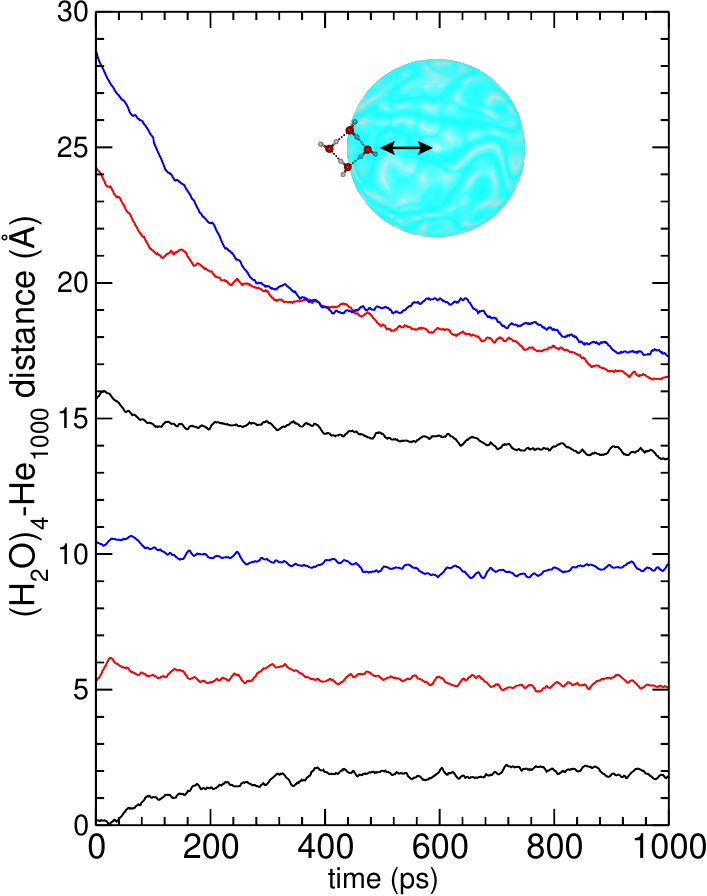}
\caption{Time evolution of the distance between the centers of mass of a water tetramer and a He$_{1000}$ helium droplet, for six different initial distances, as predicted by path-integral molecular dynamics simulations at 2~K.} 
\label{fig:PIMD}
\end{figure}

6 separating distances were chosen, at 30, 25, 15, 10, 5, and 0~\AA, respectively. Fig.~\ref{fig:PIMD} shows the variations of the distance between the water cluster and helium droplet partners, as defined from the centroids of their centers of mass.
The diffusing motion of the water cluster inside the droplet is very limited and amounts to a few \AA, unless it starts outside the droplet itself at distances greater than 20~\AA. In this case, the dynamics starts with a rather slow submersion process taking place over a few hundreds of picoseconds, with an even slower diffusion towards the droplet center with a magnitude that remains very limited after one nanosecond. This submersion process is associated with some helium evaporation occuring as a primary mechanism to release the excess energy, and which may further slow down the radial diffusion process.

\providecommand*{\mcitethebibliography}{\thebibliography}
\csname @ifundefined\endcsname{endmcitethebibliography}
{\let\endmcitethebibliography\endthebibliography}{}